\documentclass[aps,prd,superscriptaddress,twoside,twocolumn,nofootinbib,10pt,%
showpacs,floatfix]{revtex4-1}

\usepackage{epsfig}
\usepackage[dvipdfmx]{color}
\usepackage{graphicx}
\usepackage{amsmath}
\usepackage{epstopdf}
\usepackage{ulem}
\usepackage{subfigure}
\usepackage{here}

\allowdisplaybreaks

\begin{document}

\title{Magnetic field effect on nuclear matter from skyrmion crystal model}

\author{Mamiya Kawaguchi}
\email{mkawaguchi@hken.phys.nagoya-u.ac.jp}
\affiliation{ Department of Physics, Nagoya University, Nagoya 464-8602, Japan.}

\author{Yong-Liang Ma}
\email{yongliangma@jlu.edu.cn}
\affiliation{Center for Theoretical Physics and College of Physics, Jilin University, Changchun, 130012, China}

\author{Shinya Matsuzaki } 
\email{synya@hken.phys.nagoya-u.ac.jp}
\affiliation{ Department of Physics, Nagoya University, Nagoya 464-8602, Japan.}
\affiliation{Center for Theoretical Physics and College of Physics, Jilin University, Changchun, 130012, China}

\date{\today}


 \begin{abstract}
We explore magnetic field effects on the nuclear matter
based on the skrymion crystal approach for the first time.
It is found that the magnetic effect plays the role of a catalyzer for the topological phase transition (topological deformation for the skyrmion crystal configuration from the skrymion phase to half-skyrmion phase).
Furthermore, we observe that in the presence of the magnetic field, the inhomogeneous chiral condensate persists both in the skyrmion and half-skyrmion phases.
Explicitly, as the strength of magnetic field gets larger, the inhomogeneous chiral condensate in the skyrmion phase tends to be drastically localized, while in the half-skyrmion phase the inhomogeneity configuration is hardly affected.
It also turns out that a large magnetic effect in a low density region distorts the baryon shape to an elliptic form but the crystal structure is intact.
However, in a high density region,
the crystal structure is strongly effected by the strong magnetic field.
A possible correlation between the chiral inhomogeneity and the deformation of the skrymion configuration is also addressed. 
The results obtained in this paper might be realized in the deep interior of compact stars.
\end{abstract}
\maketitle

\section{Introduction}

Exploring the phase diagram of QCD under external magnetic source is an active and a significant 
 field in high energy physics relevant to the heavy ion physics, compact star and the evolution of early universe (see, e.g., Ref.~\cite{Andersen:2014xxa} for a recent review 
 references therein).
Studies of this kind have been performed using various effective theories
and models such as the chiral perturbation theory~\cite{Shushpanov:1997sf,Agasian:2001ym,Werbos:2007ym,Andersen:2012dz}
and Nambu-Jona-Lasinio model~\cite{Inagaki:2003yi,Menezes:2008qt,Boomsma:2009yk,Yu:2014xoa,Liu:2018zag,Liu:2018zag}.
In the present work, we make the first attempt to
study the magnetic effect on the nuclear matter
based on  the skyrmion crystal model.

In the skyrmion crystal approach baryons arise as the topological objects in a nonlinear mesonic theory -- skyrmions -- Skyrme model~\cite{Skyrme:1962vh} and, the nuclear matter is simulated by putting the skyrmions onto the crystal lattice 
and regarding the nuclear matter 
as the skyrmion matter~\cite{Klebanov:1985qi}. 
(In the present analysis, we specifically choose the face centered cubic (FCC) crystal.) In this approach, the nuclear matter, medium-modified hadron properties and the symmetry breaking patterns in a dense system
can be accessed in a unified way~\cite{Lee:2003aq,Lee:2003rj,Ma:2013ooa,Ma:2013ela} (for a recent review, see, e.g., Ref.\cite{Ma:2016gdd}).

In this paper, we include the magnetic field in the skyrmion crystal approach for the first time to study the magnetic effect on the nuclear matter with interests
particularly in the topological phase transition, inhomogeneous quark condensate and
the shape of single baryon (skyrmion).
What we have done and found can be summarized as follows:
\begin{itemize}
  \item The magnetic effect plays the role of a catalyzer for the topological phase transition (topological deformation for the skyrmion crystal configuration from the skrymion phase to half-skyrmion phase).
  The baryon energy per skyrmion (soliton mass) is enhanced by the presence of the magnetic field.

  \item
  Regarding the magnetic field dependence of the inhomogeneity of the chiral condensate in a medium modeled by the skyrmion crystal,
 it turns out that
  even in the presence of the magnetic field, the inhomogeneous chiral condensate persists both in the skyrmion and half-skyrmion phases.
  Interestingly enough,  as the strength of magnetic field gets larger, the inhomogeneous chiral condensate in the skyrmion phase tends to be drastically localized, while in the half-skyrmion phase the inhomogeneity configuration is almost intact. (Similar observations, regarding the deformation of inhomogeneities for the chiral condensate by magnetic effects, have been made in different models~\cite{Nishiyama:2015fba,Buballa:2015awa,Abuki:2016zpv,Abuki:2018wuv}.)

  \item
  As to the magnetic effect on the skyrmion configuration and the single baryon shape in the medium,
  it is found that a large magnetic strength in a low density region (in the skyrmion phase) makes the baryon shape to be elliptic,
  while the crystal configuration essentially holds.
 In contrast,  in a high density region (in the half-skyrmion phase),
  the crystal structure is significantly affected by the existence of a large magnetic field.

  \item
  A correlation between the chiral inhomogeneity and
  the deformation of the skrymion crystal configuration,
 can be seen through a nontrivial deformation due to a large magnetic field,
 which would be a novel indirect probe for the presence of the inhomogeneity of the chiral condensate in the half-skyrmion phase.
\end{itemize}

%
We anticipate that our findings as listed above might affect the equation of state of 
dense nuclear matter or compact stars having a strong magnetic field. 
Such characteristic features possibly emergent in dense matters/compact stars  
could be (indirectly) tested by future astronomical observations.

This paper is organized as follows: In sec.~\ref{sec:model} we introduce the basic setup in studying magnetic properties of the skyrmion crystal. With this preliminary setup at hand,
in Sec.~\ref{sec:Num} we show our numerical analysis for the magnetic dependences on the skyrmion crystal and some related phenomena such as the topological phase transition,
the inhomogeneous chiral condensates and the deformation of single baryon shape.
Sec.~\ref{sec:sum} is devoted to summary of our study and findings. Appendix A provides detailed computations regarding some prescription for the crystal configuration under the magnetic effect.

\section{The model of Skyrmion crystal in a magnetic field}

\label{sec:model}

In this section, we first give a brief summary of the basics of the skyrmion crystal model related to the present work following Ref.~\cite{Lee:2003aq}, and then
introduce the basic strategy for studying the magnetic properties of
the skymrion crystal.

\subsection{Skyrmion crystal}

We begin by considering
the the following Skyrme model Lagrangian \cite{Skyrme:1962vh}
\begin{eqnarray}
{\cal L}_{\rm Skyr} & = & \frac{f_\pi^2}{4}{\rm tr}\left[\partial_\mu U \partial^\mu U^\dagger\right] + \frac{1}{32g^2}{\rm tr}\left[U^\dagger \partial_\mu U,U^\dagger \partial_\nu U\right]^2,
\nonumber\\
\label{Lag1}
\end{eqnarray}
where $U$ is the chiral field embedding the pion fields. $ f_\pi$ is the pion decay constant  and $g$ is the dimensionless coupling constant, the Skyrme parameter.
In the skyrmion crystal approach, it is convenient to parameterize the chiral field $U$ as
\begin{eqnarray}
U=\phi_0+i\tau_a\phi_a,
\label{paraU}
\end{eqnarray}
with $a = 1,2,3$ and the unitary constraint $(\phi_0)^2+(\phi_a)^2=1$.
Note that with this parameterization \eqref{paraU},
the $\phi_\alpha \;\; (\alpha=0,1,2,3)$ can be rephrased in terms of quark bilinear configurations as
\begin{eqnarray}
\phi_0&\sim &\bar q q , \nonumber\\
\phi_a&\sim&\bar q i \gamma_5\tau_a q.
\label{chi_order}
\end{eqnarray}
For the later convenience, we further introduce the unnormalized fields $\bar \phi_\alpha$ which are related to normalized fields $\phi_\alpha$ through
\begin{eqnarray}
\phi_\alpha=\frac{\bar\phi_\alpha}{\sqrt{\sum_{\beta=0}^3\bar\phi_\beta\bar\phi_\beta}}.
\end{eqnarray}

For a crystal lattice with size $2L$, one can parameterize the the unnormalized field $\bar\phi_\alpha$ in terms of the Fourier series as~\cite{Lee:2003aq}
\begin{eqnarray}
\bar \phi_0(x,y,z)&=&\sum_{a,b,c}\bar \beta_{abc}\cos(a\pi x/L)\cos(b\pi y/L)\cos(c\pi z/L),\nonumber\\
\bar \phi_1(x,y,z)&=&\sum_{h,k,l}\bar \alpha_{hkl}^{(1)}\sin(h\pi x/L)\cos(k\pi y/L)\cos(l\pi z/L),\nonumber\\
\bar \phi_2(x,y,z)&=&\sum_{h,k,l}\bar \alpha_{hkl}^{(2)}\cos(l\pi x/L)\sin(h\pi y/L)\cos(k\pi z/L),\nonumber\\
\bar \phi_3(x,y,z)&=&\sum_{h,k,l}\bar \alpha_{hkl}^{(3)}\cos(k\pi x/L)\cos(l\pi y/L)\sin(h\pi z/L),
\nonumber\\
\label{ansatz_1}
\end{eqnarray}
where $\bar{\alpha}$ and $\bar{\beta}$ are free parameters which are determined by minimizing the energy of the system.

For a particular crystal lattice, the Fourier coefficients $\bar{\alpha}$ and $\bar{\beta}$ are not independent of each other. For example,  the FCC, which will be used in the present work, possesses
the following symmetry structure in position space and corresponding isospin space:
\begin{itemize}
  \item Reflection symmetry:

   In position space $(x,y,z)\leftrightarrow (-x,y,z)$ and in isospin space $ (\phi_0,\phi_1,\phi_2,\phi_3)\leftrightarrow (\phi_0,-\phi_1,\phi_2,\phi_3)$;

  \item Three-fold  symmetry:

  In position space $(x,y,z)\leftrightarrow (z,x,y)$ and in isospin space $ (\phi_0,\phi_1,\phi_2,\phi_3)\leftrightarrow (\phi_0,\phi_3,\phi_1,\phi_2)$;

  \item Four-fold  symmetry:

  In position space $ (x,y,z)\leftrightarrow (y,-x,z)$ and in isospin space $
 (\phi_0,\phi_1,\phi_2,\phi_3)\leftrightarrow (\phi_0,\phi_2,-\phi_1,\phi_3)$;

  \item Translation symmetry:

  In position space $(x,y,z)\leftrightarrow (x+L,y+L,z)$
   and in isospin space $ (\phi_0,\phi_1,\phi_2,\phi_3)\leftrightarrow (\phi_0,-\phi_1,-\phi_2,\phi_3)$.
\end{itemize}
Hence  the Fourier coefficients $\bar\beta_{abc},
\bar\alpha_{hkl}^{(1,2,3)}$ have the following relations~\cite{Lee:2003aq}:
\begin{itemize}
  \item From three-fold  symmetry:

  $\bar \beta_{abc}=\bar\beta_{bca}=\bar \beta_{cab},\;\;
\bar \alpha_{hkl}^{(1)}=\bar \alpha_{hkl}^{(2)}=\bar \alpha_{hkl}^{(3)}$;

  \item From four-fold symmetry:

  $\bar \beta_{abc}=\bar \beta_{acb}=\bar \beta_{cba}=\bar \beta_{bac} \;\; \bar\alpha_{hkl}=\bar\alpha_{hlk}$;

  \item From translation symmetry:

  $a, b, c$ are all even numbers or odd numbers, and if $h$ is even, then $k, l$ are restricted to odd numbers,
otherwise even numbers.
\end{itemize}


\subsection{Introducing magnetic field in skyrmion crystal }

Next, let us discuss how to introduce the magnetic field in the skyrmion crystal model.

The magnetic field can be incorporated into the Skyrme model \eqref{Lag1} by replacing the derivative operator with the gauge covariant one,
\begin{eqnarray}
D_\mu U & = & \partial_\mu U-ieA_\mu  [Q_E,U],
\end{eqnarray}
where $e$ is the electromagnetic coupling constant
and $Q_E=\frac{1}{6}+\frac{1}{2}\tau_3$ is the electric charge matrix
with $\tau_3$ being the third component of Pauli matrix.
In the present work, we consider a constant magnetic field $B$ along the $z$ direction. 
Then, 
taking into account the residual $O(2)$ symmetry for the $x$-$y$ plane perpendicular to the $B$-axis, we choose the symmetric gauge
\begin{eqnarray}
A_\mu & = & {} -\frac{1}{2}By\delta_\mu^{\;\;1}+\frac{1}{2}Bx\delta_\mu^{\;\;2}.
\end{eqnarray}

In terms of the parameterization \eqref{paraU}, we can write the covariant derivative operator as
\begin{eqnarray}
D_\mu U & = & \partial_\mu\phi_0 + iD_\mu\phi_1\tau_1 + iD_\mu\phi_2\tau_2 + i\partial_\mu\phi_3\tau_3,
\end{eqnarray}
where $D_\mu \phi_1=\partial_\mu \phi_1-eA_\mu\phi_2$ and $D_\mu \phi_2=\partial_\mu \phi_2+eA_\mu\phi_1$.
It should be noted that the translational symmetry in the skyrmion crystal might be incompatible with the gauge covariance after fixing the gauge.
In the symmetric gauge~\footnote{Since we model the nuclear matter by the skyrmion crystal, it is necessary to choose the symmetric gauge which keeps the $O(2)$ symmetry appropriate to form the crystal. This kind of gauge dependence can also be seen in the ladder approximations, where as is well known, QCD observables are modeled and computed with some specific gauge choice to be consistent with the associated chiral symmetry
(e.g., see~\cite{Maskawa:1975hx,Kugo:1992pr,Kugo:1992zg,Bando:1993qy}).},
for instance, under a translation in the position space,  $(x,y,z)\to (x+L,y+L,z)$, the covariant derivative $D_x\phi_1$ goes like
\begin{eqnarray}
\partial_x\phi_1(x,y,z)& \to & {}  -\partial_x\phi_1(x,y,z)
\,, \nonumber\\
 y\phi_2(x,y,z)& \to &{} -(y+L)\phi_2(x,y,z),
\end{eqnarray}
which obviously does not reflect the gauge covariance.
However, even if the magnetic field is present, the translational invariance should be preserved in the skyrmion crystal.

To settle down the issue on the compatibility
of the translational invariance with the gauge covariance,
we go back to the step for the discretization of $y\bar\phi_2(x,y,z)$ with respect to the Fourier momenta:
\begin{widetext}
\begin{eqnarray}
y\bar\phi_2(x,y,z)
&=&
y\int_{-\infty}^{\infty} \frac{dp_x}{(2\pi)}\int_{-\infty}^{\infty} \frac{dp_y}{(2\pi)}\int_{-\infty}^{\infty} \frac{dp_z}{(2\pi)}
\bar\phi_2(p_x,p_y,p_z)
e^{i{\vec p}\cdot {\vec x}}
\nonumber\\
&=&
\int_{0}^{\infty} \frac{dp_x}{(2\pi)}\int_{0}^{\infty} \frac{dp_y}{(2\pi)}\int_{0}^{\infty} \frac{dp_z}{(2\pi)}
\bar\phi_2(p_x,p_y,p_z)
8i\cos(p_xx)\left(-\partial_{p_y}\cos(p_yy)\right)\cos(p_zz)\nonumber\\
&\xrightarrow{{\rm discretization}}&
-\sum_{h,k,l}\bar\alpha_{h,k,l}^{ (2)}
\cos(l\pi x/L)
\frac{\cos\{(h+2)\pi y/L\}-\cos(h\pi y/L)}
{2\pi/L}
\cos(k\pi z/L)\nonumber\\
&\equiv&[y\bar\phi_2]_{\rm disc}(x,y,z),
\label{disc1}
\end{eqnarray}
where the second equality  was obtained by the reflection symmetry.
Now we make the translation in the position space $(x,y,z)\to (x+L,y+L,z)$, then
 $[y\bar\phi_2]_{\rm disc}(x,y,z)$ transforms like
\begin{eqnarray}
[y\bar\phi_2]_{\rm disc}(x,y,z)&\to &{} -[y\bar\phi_2]_{\rm disc}(x,y,z).
\end{eqnarray}
Thus, when we use $[y\bar\phi_2]_{\rm disc}$ instead of $y\bar\phi_2$, the covariant derivative $D_x\phi_1$ keeps the gauge covariance after the translation.
Similarly, one can get the following discritization forms
\begin{eqnarray}
{[}x\bar\phi_1]_{\rm disc}(x,y,z)&=&
-\sum_{h,k,l}\bar\alpha_{h,k,l}
\frac{\cos\{(h+2)\pi x/L\}-\cos(h\pi x/L)}{2\pi/L}
\cos(k\pi y/L)\cos(l\pi z/L)
\,, \nonumber\\
{[}y\bar\phi_1]_{\rm disc}(x,y,z)
&=&
\sum_{h,k,l}\bar\alpha_{h,k,l}
\sin(h\pi x/L)
\frac{\sin\{(k+2)\pi y/L\}-\sin(k\pi y/L)}
{2\pi/L}
\cos(l\pi z/L)
\,, \nonumber\\
{[}x\bar\phi_2]_{\rm disc}( x,y,z)
&=&
\sum_{h,k,l}\bar\alpha_{h,k,l}
\frac{\sin\{(l+2)\pi x/L\}-\sin(l\pi x/L)}
{2\pi/L}
\sin(h\pi y/L)\cos(k\pi z/L).
\label{disc2}
\end{eqnarray}

With the prescribed discretization method at hand, let us check whether the other symmetries in the FCC are kept in the presence of a magnetic field.
For instance, the covariant derivative of $\phi_1$ is transformed as
\begin{eqnarray}
{[}D_i\phi_{1}]_{\rm disc}(x,y,z)&\xrightarrow{\mbox{ reflection: }(x,y,z)\to(-x,y,z)}&
\begin{cases}
(i=x)\;\;{[}D_x\phi_{1}]_{\rm disc}(x,y,z)\\
(i= y)\;\;{}-{[}D_y\phi_{1}]_{\rm disc}(x,y,z)\\
(i= z)\;\;{}-{[}D_z\phi_{1}]_{\rm disc}(x,y,z)
\end{cases}
\,, \nonumber\\
{[}D_i\phi_{1}]_{\rm disc}(x,y,z)&\xrightarrow{\mbox{ translation:}(x,y,z)\to (x+L,y+L,z)}&
{} -{[}D_i\phi_{1}]_{\rm disc}(x,y,z)
\,, \nonumber\\
{[}D_i\phi_{1}]_{\rm disc}(x,y,z)&\xrightarrow{\mbox{ two-fold  for z axis:}(x,y,z)\to (y,-x,z)}&
\begin{cases}
(i=x)\;\;
(\partial_y \phi_2+eA_y\phi_1)_{\rm disc}(x,y,z)
\\
(i=y)\;\;
{} -(\partial_x \phi_2+eA_x\phi_1)_{\rm disc}(x,y,z)
\\
(i=z)\;\;
{[}D_z\phi_{2}]_{\rm disc}(x,y,z)
\end{cases}
\,, \nonumber\\
{[}D_i\phi_{1}]_{\rm disc}(x,y,z)&\xrightarrow{\mbox{ two-fold  for x axis}:(x,y,z)\to (x,z,-y)}&
\begin{cases}
(i=x)\;\;
(\partial_x \phi_1-eA_x\phi_3)_{\rm disc}(x,y,z)
\\
(i=y)\;\;
(\partial_z \phi_1-eA_z\phi_3)_{\rm disc}(x,y,z)
\\
(i=z)\;\;
-(\partial_y \phi_1-eA_y\phi_3)_{\rm disc}(x,y,z)
\end{cases}
\,, \nonumber\\
{[}D_i\phi_{1}]_{\rm disc}(x,y,z)&\xrightarrow{\mbox{ three-fold }:(x,y,z)\to (z,x,y)}&
\begin{cases}
(i=x)\;\;
(\partial_z \phi_3-eA_z\phi_1)_{\rm disc}(x,y,z)
\\
(i=y)\;\;
(\partial_x \phi_3-eA_x\phi_1)_{\rm disc}(x,y,z)
\\
(i=z)\;\;
(\partial_y \phi_3-eA_y\phi_1)_{\rm disc}(x,y,z)
\end{cases}
.
\end{eqnarray}
\end{widetext}
One can see easily that, as naively expected, the two-fold  symmetry for $x(y)$ axis and the three-fold symmetry are explicitly broken by the magnetic field in $z$-direction.

\subsection{Baryon number density of skyrmion crystal in a magnetic field}

In the presence of a magnetic field, the baryon number  current is evaluated as~\cite{He:2015zca}
\begin{eqnarray}
j_{B}^\mu & = &j^\mu_{W}+{\cal J}_{eB}^\mu,
\end{eqnarray}
where
\begin{eqnarray}
j^\mu_W & = & \frac{1}{24\pi^2}\epsilon^{\mu\nu\rho\sigma}{\rm tr}
\left[
(\partial_\nu U\cdot U^\dagger)(\partial_\rho U\cdot U^\dagger)(\partial_\sigma U\cdot U^\dagger)
\right]
\,, \nonumber\\
{\cal J}_{eB}^\mu & = & \frac{1}{16\pi^2}\epsilon^{\mu\nu\rho\sigma}
\nonumber\\
& &{} \times {\rm tr}\left[ie(\partial_\nu {A}_\rho) Q_E
(\partial_\sigma U\cdot U^\dagger +U^\dagger\partial_\sigma U) \right. \nonumber\\
& &\left. \qquad\; {} +ie{A}_\nu Q_E(\partial_\rho U \partial_\sigma U^\dagger-\partial_\rho U^\dagger \partial_\sigma U)
\right],
\label{current}
\end{eqnarray}
in which $j^\mu_{W}$ corresponds to the winding number current and ${\cal J}_{eB}^\mu$ is the induced current by the magnetic field. Under the symmetric gauge, the time component of the induced current takes the form
\begin{eqnarray}
{\cal J}_{eB}^0 & = &{} - \frac{eB}{4\pi^2}\left[(\partial_z\phi_3)\phi_0-(\partial_z\phi_0)\phi_3 \right]\nonumber\\
& &{} + \frac{eB}{8\pi^2} \left[\left\{
[y\partial_y\phi_3]_{\rm disc}(\partial_z\phi_0)-[y\partial_y\phi_0]_{\rm disc}(\partial_z\phi_3)\right\} \right. \nonumber\\
& & \qquad\quad \left. {} -
\left\{
(\partial_z\phi_3)[x\partial_x\phi_0]_{\rm disc}-(\partial_z\phi_0)[x\partial_x\phi_3]_{\rm disc}
\right\}\right].
\nonumber\\
\label{app}
\end{eqnarray}
Explicit expressions of the discretized form involving a derivative
as above are given in Appendix A.
Then, the total baryon number density $\rho_B(x,y,z)$ is written as
\begin{eqnarray}
\rho_B(x,y,z)&=&j_W^0(x,y,z)+{\cal J}^0_{eB}(x,y,z)\nonumber\\
&\equiv&\rho_W(x,y,z)+\tilde\rho_{eB}(x,y,z),
\label{bar_dens_eq}
\end{eqnarray}
where $\rho_W(z,y,z)$ is the winding number density and $\tilde\rho_{eB}(x,y,z)$ is the baryon number density induced by a magnetic field. The baryon number is obtained by performing the spacial integration $N_B=\int_{\rm cube}d^3 x \rho_B$.
As in the case without a magnetic field,
the baryon number in a single FCC crystal is normalized as
\begin{eqnarray}
N_B=\int_{\rm cube}d^3 x \rho_B=4,
\end{eqnarray}
because $\int_{\rm cube}\tilde\rho_{eB}=0$~\footnote{By using
 Eqs.(\ref{ansatz_1}) and the symmetry relations, one can derive
\begin{eqnarray}
& & \int_{\rm cube}d^3x\,(\partial_z\phi_3)\phi_0=
\int_{\rm cube}d^3x\,\phi_3\partial_z\phi_0 = 0\,,
\nonumber\\
& & \int_{\rm cube}d^3x\,\left[y(\partial_y\phi_3)\right]_{\rm disc}(\partial_z\phi_0)=
\int_{\rm cube}d^3x\,\left[y(\partial_y\phi_0)\right]_{\rm disc}(\partial_z\phi_3)
 = 0
\,, \nonumber\\
& & \int_{\rm cube}d^3x\,\left[x(\partial_x\phi_3)\right]_{\rm disc}(\partial_z\phi_0)=
\int_{\rm cube}d^3x\,\left[x(\partial_x\phi_0)\right]_{\rm disc}(\partial_z\phi_3)
= 0 \,.
\nonumber
\end{eqnarray}
Here,  we used the same Fourier expansion form for  $\phi_\alpha$ as that for  $\bar\phi_\alpha$,
as in Eq.(\ref{ansatz_1}).
Thus one can find
 $\int_{\rm cube}{\cal J}_B^0(eB)=0$, hence
the baryon number is conserved under a magnetic field.
\label{fot1}
}
.

\section{Magnetic field effect in skyrmion crystal}

\label{sec:Num}

With the above setup, we are now ready to numerically simulate the magnetic effect on the nuclear matter. For this purpose, we  take the following typical values~\cite{Ma:2016npf}
\begin{eqnarray}
f_\pi & = & 92.4~{\rm MeV}, \qquad g=5.93.
\end{eqnarray}


\subsection{Per-baryon energy $E/N_B$}\label{per_E}

The per-baryon energy is given by
\begin{eqnarray}
E/N_B & = &{} \frac{1}{4}\int_{\rm cube}d^3x {\cal H}_{\rm Skyr}^{B},
\label{eq:BE}
\end{eqnarray}
with ${\cal H}_{\rm Skyr}^{B}$ being the static skyrmion energy in the presence of the external magnetic field introduced as above.
In \eqref{eq:BE}, the per-baryon energy is given as
a function of the Fourier coefficients $\bar\beta_{abc},\bar \alpha_{hkl}^{(i)}$, the crystal size $L$ and a magnetic field strength $eB$. For a chosen crystal size $L$ and a magnetic strength $eB$, we use the Fourier coefficients as variational parameters to minimize the per-baryon energy.
In this way, the per-baryon energy can be calculated as a function of crystal size $L$ and a magnetic field strength $eB$.

\begin{figure}[h]
 \begin{center}
   \includegraphics[width=8cm]{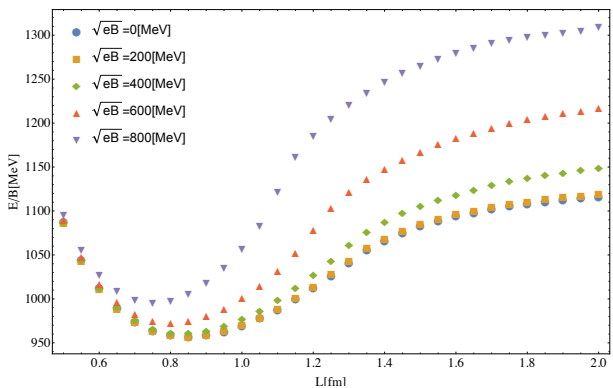}
  \end{center}
 \caption{The per-baryon energy in magnetic field as a function of the crystal size $L$. }
\label{energy_sym}
\end{figure}

In Fig.~\ref{energy_sym} we plot the per-baryon energy as a function of the crystal size with typical values for the magnetic field strengths fixed.
Note first that the bottommost curve in Fig.~\ref{energy_sym}
precisely reproduces the crystal-size dependence of the per-baryon energy in Ref.~\cite{Ma:2016npf} without the magnetic field $(eB=0)$, which manifests a check of our numeric code.
From Fig.~\ref{energy_sym} we see that, for a fixed crystal size $L$, the per-baryon energy monotonically increases as the magnetic field grows up.
In the low density region (large crystal size $L$),
this tendency can be compared with the result obtained in \cite{He:2015zca}
for an isolated skyrmion (not in the skyrmion crystal).
One can also see an interesting discrepancy in comparison with the reference:
the skyrmion energy analyzed in~\cite{He:2015zca} has a minimum with respect to $eB$, due to a destructive interference between the ${\cal O}(eB)$ and ${\cal O}\bigl((eB)^2\bigl)$ terms in the per-baryon energy functional. On the contrary, this is not the case for our present work. In the skyrmion crystal approach, the ${\cal O} (eB)$ terms disappear when the per-baryon energy functional is integrated over the volume of the crystal, which is
due to the crystal structure symmetry~\footnote{
For example,
the Lagrangian has ${\cal O}(eB)$ term such as
$eA_i\phi_2\partial_i\phi_1=eA_x\phi_2\partial_x\phi_1+eA_y\phi_2\partial_y\phi_1$.
By using (\ref{ansatz_1}) and symmetry relations, one finds
\begin{eqnarray}
\int d^3x eA_x\phi_2\partial_x\phi_1&=&-\frac{eB}{2}\int d^3x\,[ y \phi_2]_{\rm disc}\partial_x\phi_1
= 0.\notag
\end{eqnarray}
Also, the integration for $eA_y\phi_2\partial_y\phi_1$ becomes  0 in a way similar to footnote\ref{fot1}.}, so that only ${\cal O}\bigl((eB)^2\bigl)$ terms survive.
This observation is supported from the magnetic field dependence of the per-baryon energy at low density in Fig.~\ref{energy_sym}.

Unlike the low density region, in the high density region (small $L$), the per-baryon energy hardly gets affected by the magnetic field. To understand this tendency we rescale the position space by the crystal size $x=x'/L$ and rewrite the covariant derivative as $D_x\phi_1=(\partial_{x'}\phi_1+\frac{eB\,L^2}{2}y'\phi_2)/L$. This shows that the magnetic effect becomes weaker when the crystal size is reduced.


\subsection{Topological phase transition and its related phenomena: inhomogeneity of chiral condensate}

\label{phi0}

In the skyrmion approach to a nuclear matter, a novel phenomena, which is not observed in other approaches,
is the so-called skyrmion to half-skyrmion transition,
where the FCC crystal with one skyrmion (baryon-number $1$)
at each vertex factorizes to cubic-centered (CC) crystal
with a half-skymrion (having baryon-number $1/2$)
at each crystal vertex~\cite{Kugler:1988mu,Kugler:1989uc}.
The order parameter of this transition is the space-averaged $\phi_0$,
\begin{eqnarray}
\langle\phi_0\rangle & = &{} \frac{1}{(2L)^3}\int_0^{2L} d^3x \phi_0.
\end{eqnarray}
We will call this phenomenon "topological phase transition", 
although there is no paradigmatic phase transition.
Hereafter we refer to the skyrmion crystal of the FCC as the skyrmion phase
and the CC crystal of the half-skyrmion as the half-skyrmion phase.

\begin{figure}[H]
 \begin{center}
   \includegraphics[width=8cm]{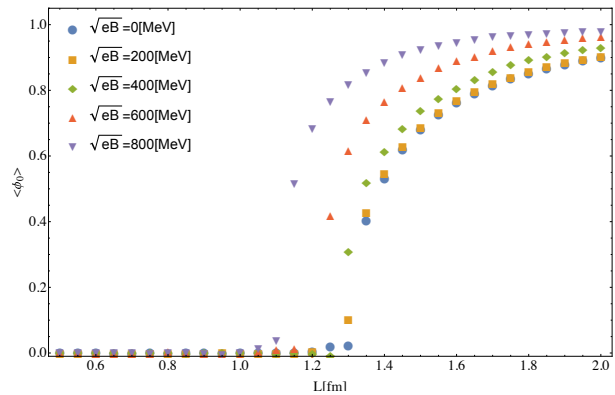}
  \end{center}
 \caption{ The order parameter for the topological phase transition,
 $\langle\phi_0\rangle$, in magnetic field, as a function of the crystal size $L$. }
\label{vev_sym}
\end{figure}

Fig.~\ref{vev_sym} shows the magnetic-field dependence on the order parameter
for the topological phase transition, $\langle \phi_0\rangle$.
Again, without a magnetic field  ($\sqrt{eB}=0$),
the critical point at $\langle \phi_0\rangle=0$ agrees
with that obtained in~\cite{Ma:2016npf} (see the bottommost curve in the figure).
From Fig.~\ref{vev_sym},
one can see that as the magnetic field increases,
the phase transition point is shifted to a high density region
and the value of the order parameter gets larger. This can be regarded as the magnetic catalysis for the topological phase transition.

\begin{figure*}[!htpb]
\centering
{
\begin{minipage}[b]{0.3\textwidth}
\includegraphics[width=1\textwidth]{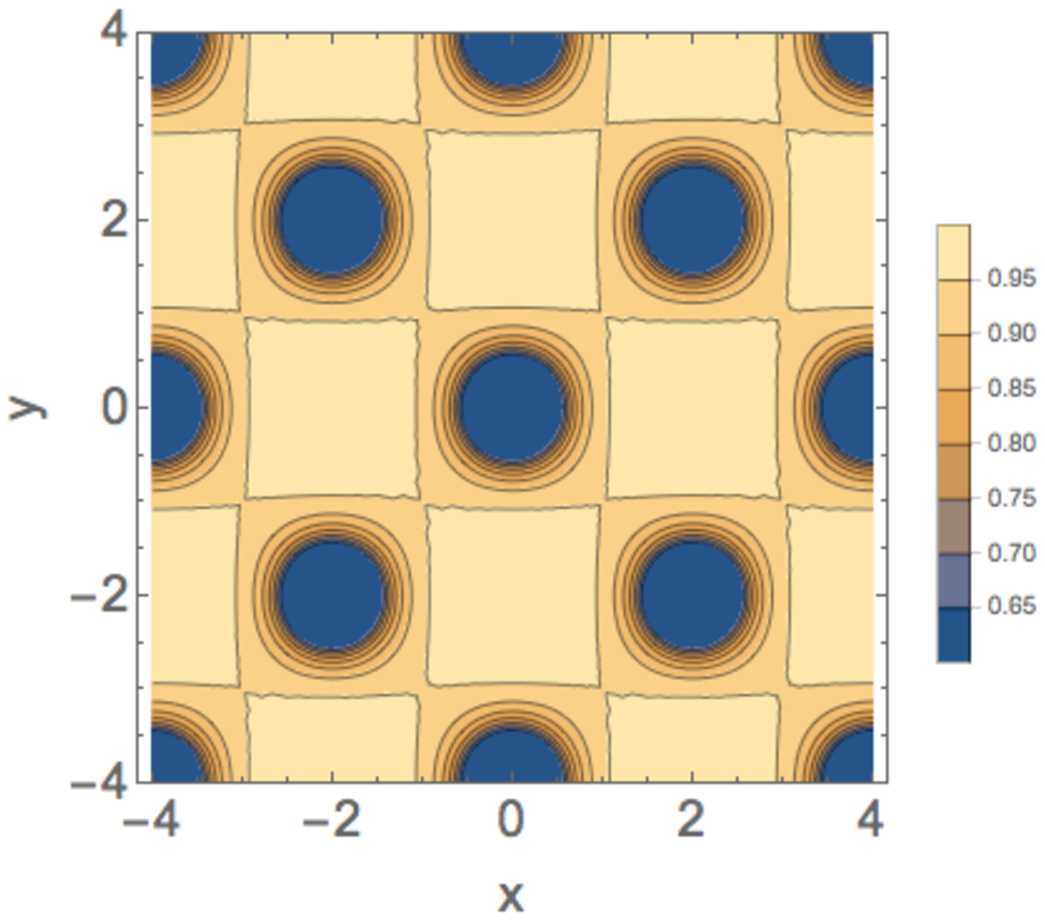}
\end{minipage}
}
{
\begin{minipage}[b]{0.3\textwidth}
\includegraphics[width=1\textwidth]{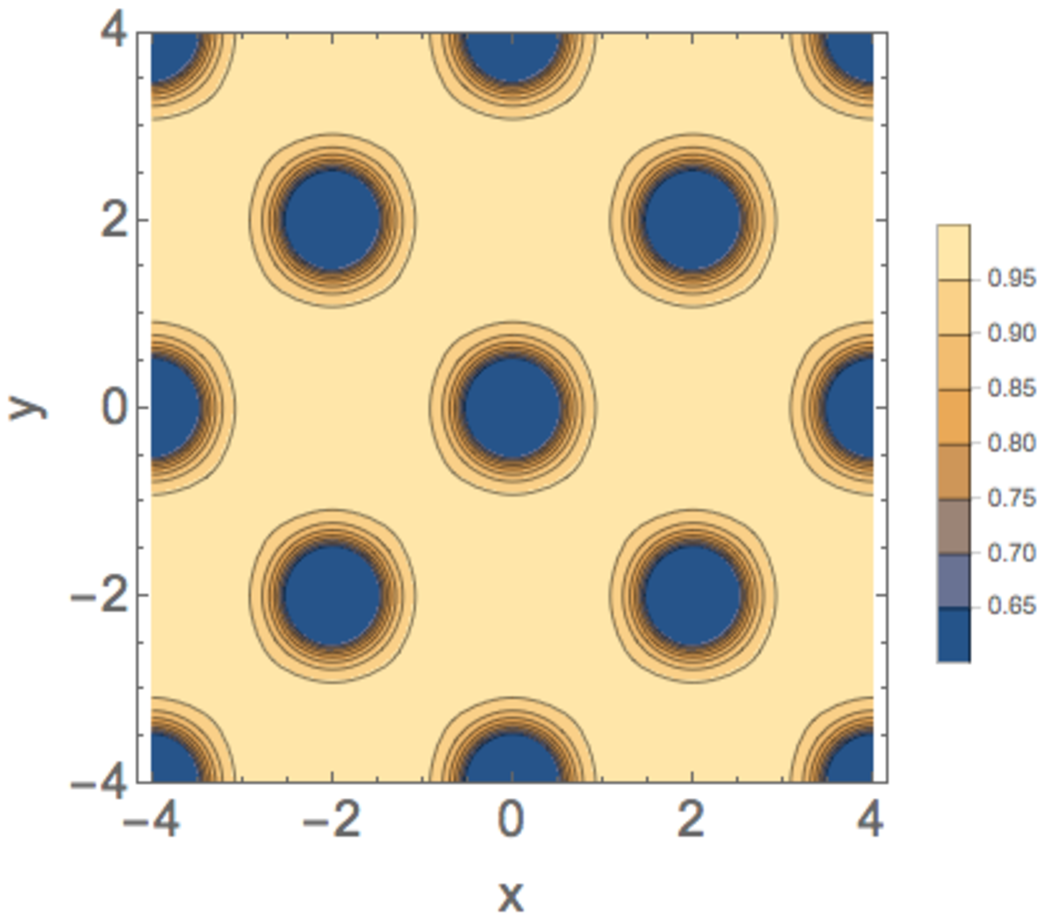}
\end{minipage}
}
{
\begin{minipage}[b]{0.3\textwidth}
\includegraphics[width=1\textwidth]{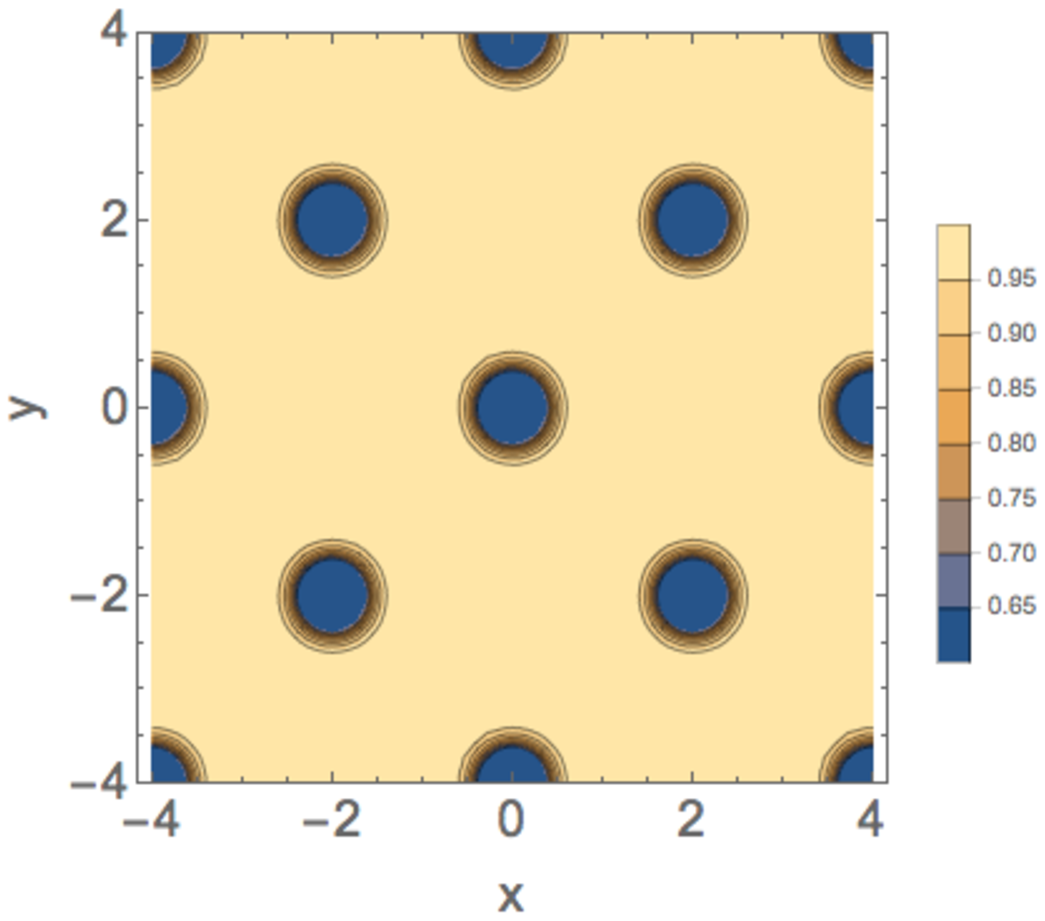}
\end{minipage}
}
\caption[]{The distributions of $\phi_0(x,y,0)$ at $L=2.0~{\rm fm}$ (in skyrmion phase) for $\sqrt{eB}=0$ (left),   $\sqrt{eB}=400~{\rm MeV}$(middle) and $\sqrt{eB}=800~{\rm MeV}$(right).}
\label{dphi0l}
\end{figure*}

\begin{figure*}[!htpb]
\centering
{
\begin{minipage}[b]{0.3\textwidth}
\includegraphics[width=1\textwidth]{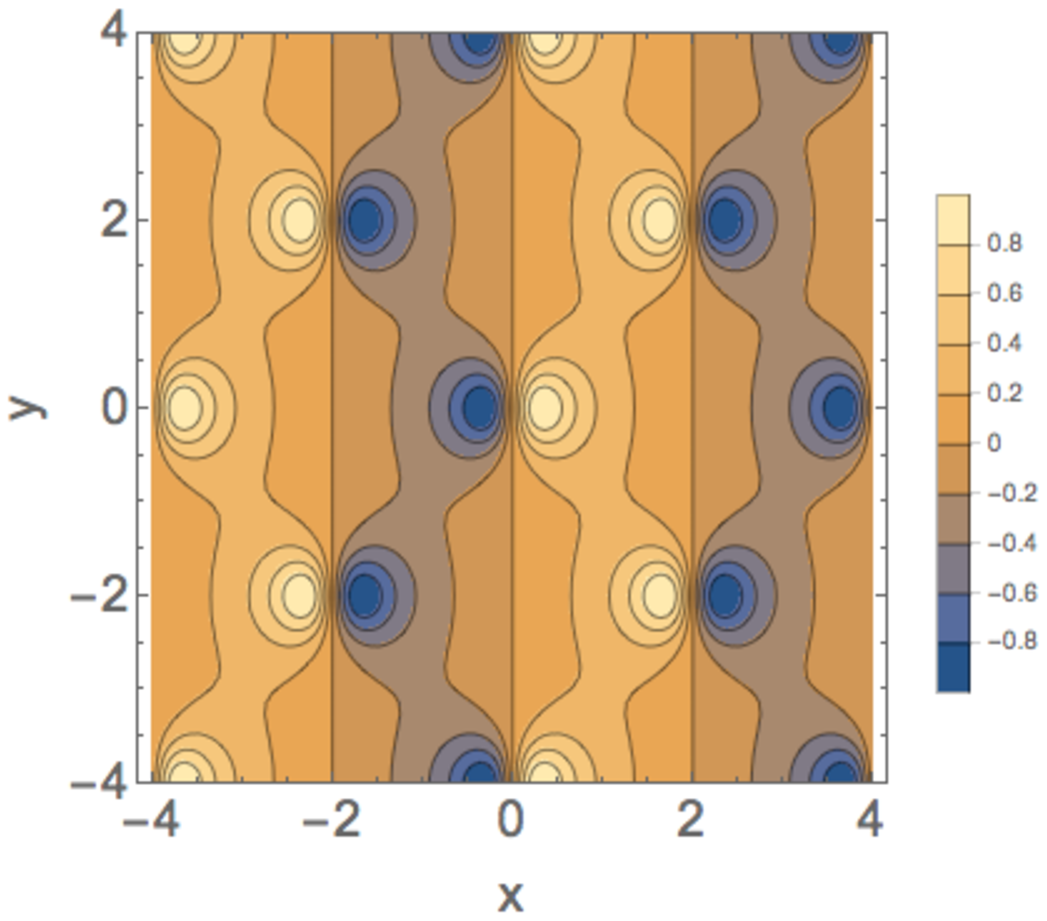}
\end{minipage}
}
{
\begin{minipage}[b]{0.3\textwidth}
\includegraphics[width=1\textwidth]{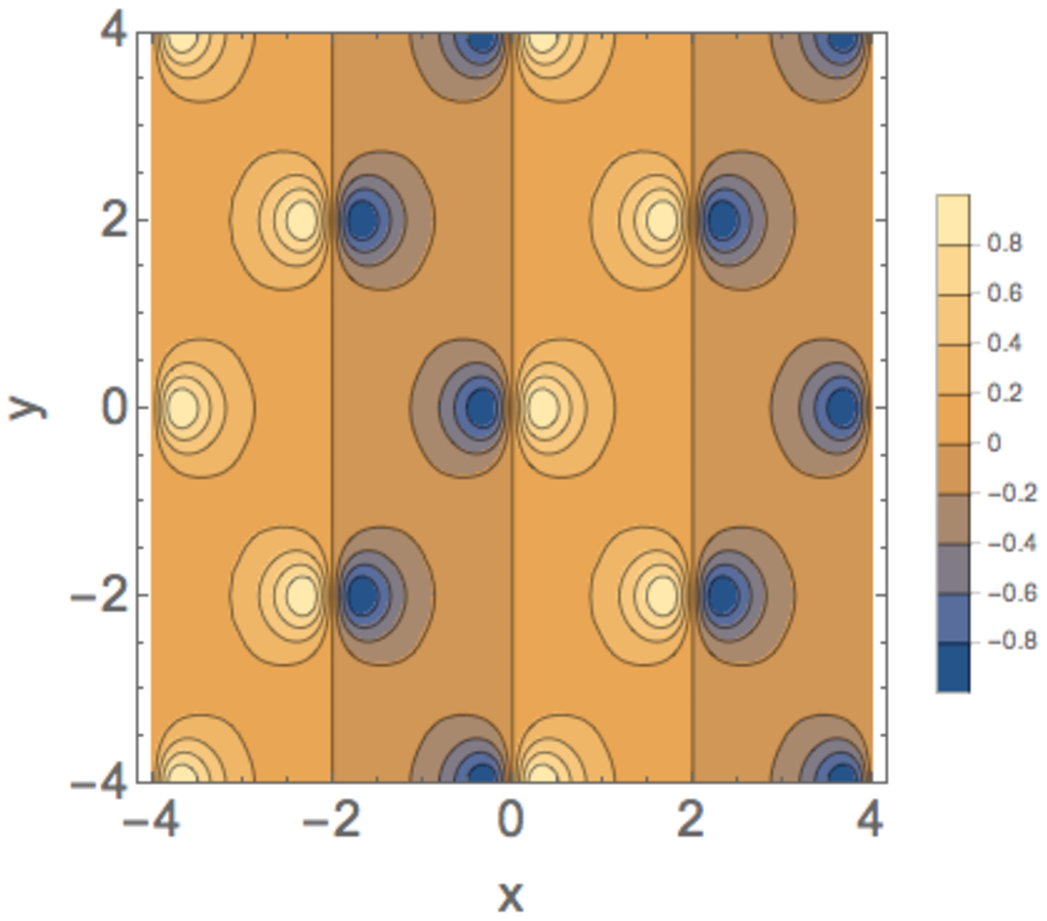}
\end{minipage}
}
{
\begin{minipage}[b]{0.3\textwidth}
\includegraphics[width=1\textwidth]{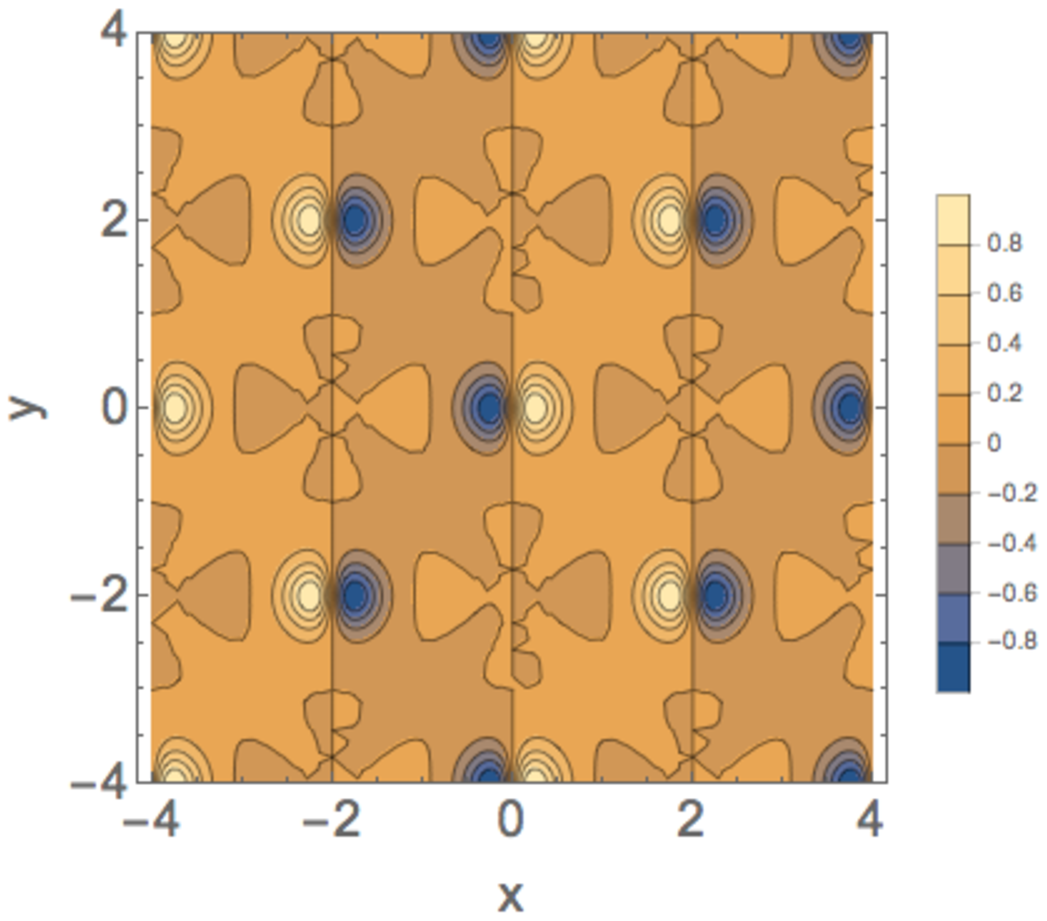}
\end{minipage}
}
\caption[]{The distributions of $\phi_1(x,y,0)$ at $L=2.0~{\rm fm}$ (in skyrmion phase) for $\sqrt{eB}=0$ (left),  $\sqrt{eB}=400~{\rm MeV}$(middle) and $\sqrt{eB}=800~{\rm MeV}$(right).
 }
  \label{dphi1l}
\end{figure*}

\begin{figure*}[!htpb]
\centering
{
\begin{minipage}[b]{0.4\textwidth}
\includegraphics[width=1\textwidth]{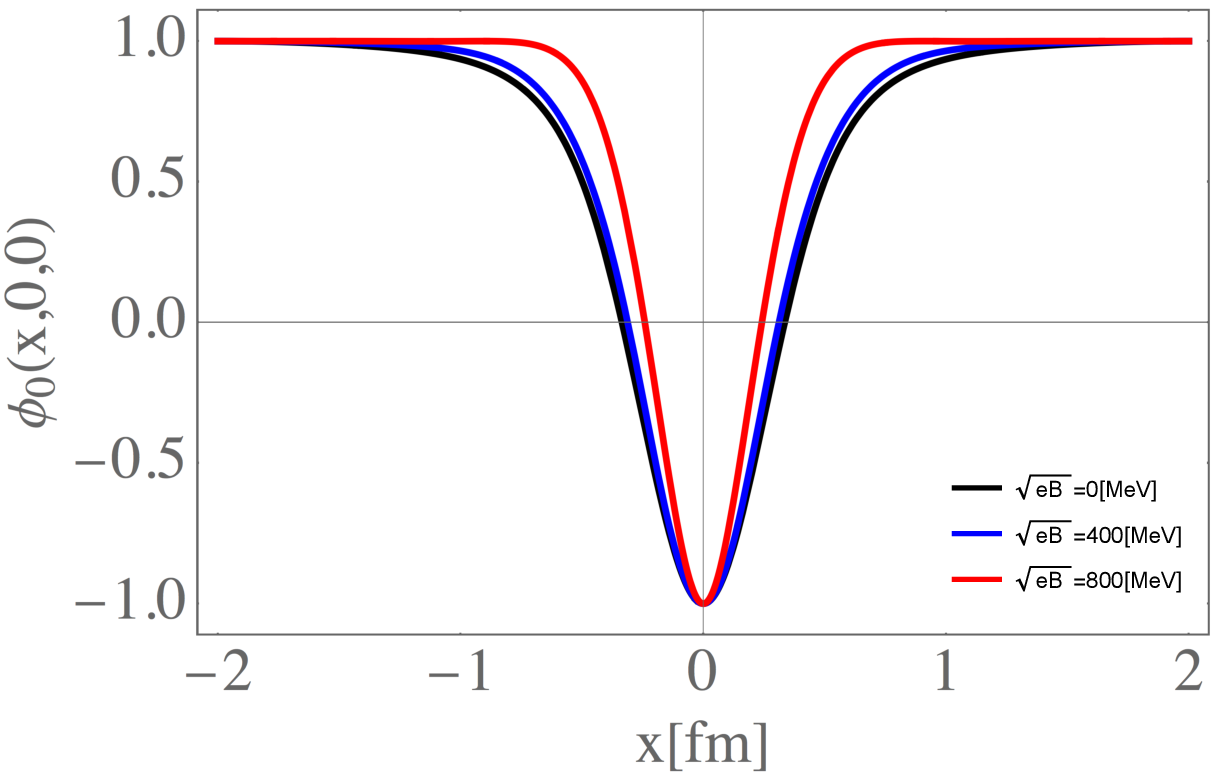}
\end{minipage}
}
{
\begin{minipage}[b]{0.4\textwidth}
\includegraphics[width=1\textwidth]{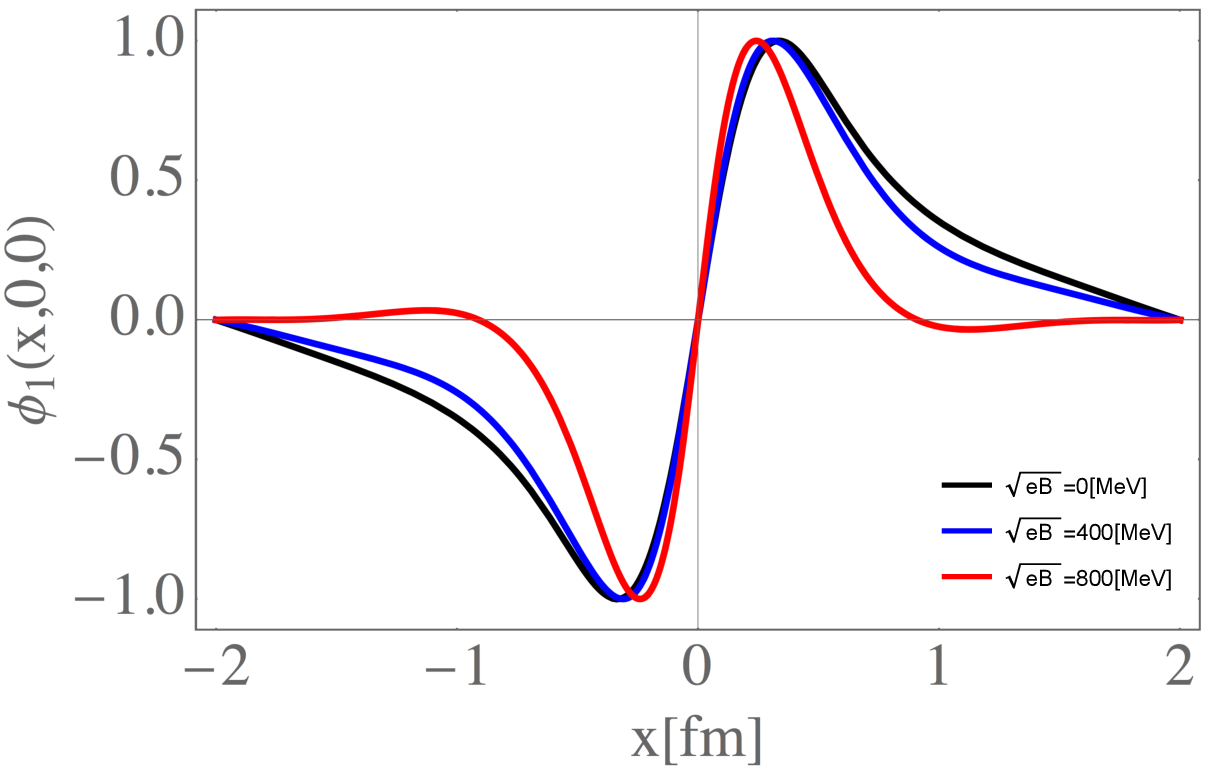}
\end{minipage}
}
\caption[]{The distributions of $\phi_0(x,0,0)$ (left panel) and $\phi_1(x,0,0)$
  (right panel) at $L=2.0~{\rm fm}~$ (in skyrmion phase) with $\sqrt{eB}$ varied.
 }
  \label{dphi-x-S}
\end{figure*}

In addition to the topological phase transition, the spatial distribution of the $\phi_0 $ can actually be thought of as
an inhomogeneous chiral condensate with $\phi_a$ as its chiral partner. We plot in Figs.~\ref{dphi0l} and~\ref{dphi1l} the magnetic dependence of the distributions of $\phi_0(x,y,z)$ and $\phi_1(x,y,z)$
at $L=2.0~{\rm fm}$ (in the skyrmion phase). The left panels in these figures show the inhomogeneities of $\phi_0$ and $\phi_1$ in the absence of the magnetic field, where the inhomogeneous configurations take the form like a ``pulse'' for the $\phi_0 (\sim \bar{q}q)$ and a ``wave'' for the $\phi_1 (\sim \bar{q} i \gamma_5 \tau^1 q)$. These two panels agree with the analysis in~\cite{Harada:2015lma}. Turning on the magnetic field (middle and right panels), one notices a striking phenomenon: as $eB$ gets bigger, the $\phi_0$ and $\phi_1$ inhomogeneities tend to be drastically localized at the vertices of the crystal (keeping each shape of the ``pulse''- and ``wave-'' like form). And, because of the dramatic localization, the density $n_{1/2}$ at which the skyrmion matter transits to half-skyrmion matter becomes larger. This novel tendency can more easily be captured by zooming in the $y=z=0$ plane, as depicted in Fig.~\ref{dphi-x-S}. Similar observations, regarding the deformation of inhomogeneities for the chiral condensate by magnetic effects, have been made in different models~\cite{Nishiyama:2015fba,Buballa:2015awa,Abuki:2016zpv,Abuki:2018wuv}.

\begin{figure*}[!htpb]
\centering
{
\begin{minipage}[b]{0.3\textwidth}
\includegraphics[width=1\textwidth]{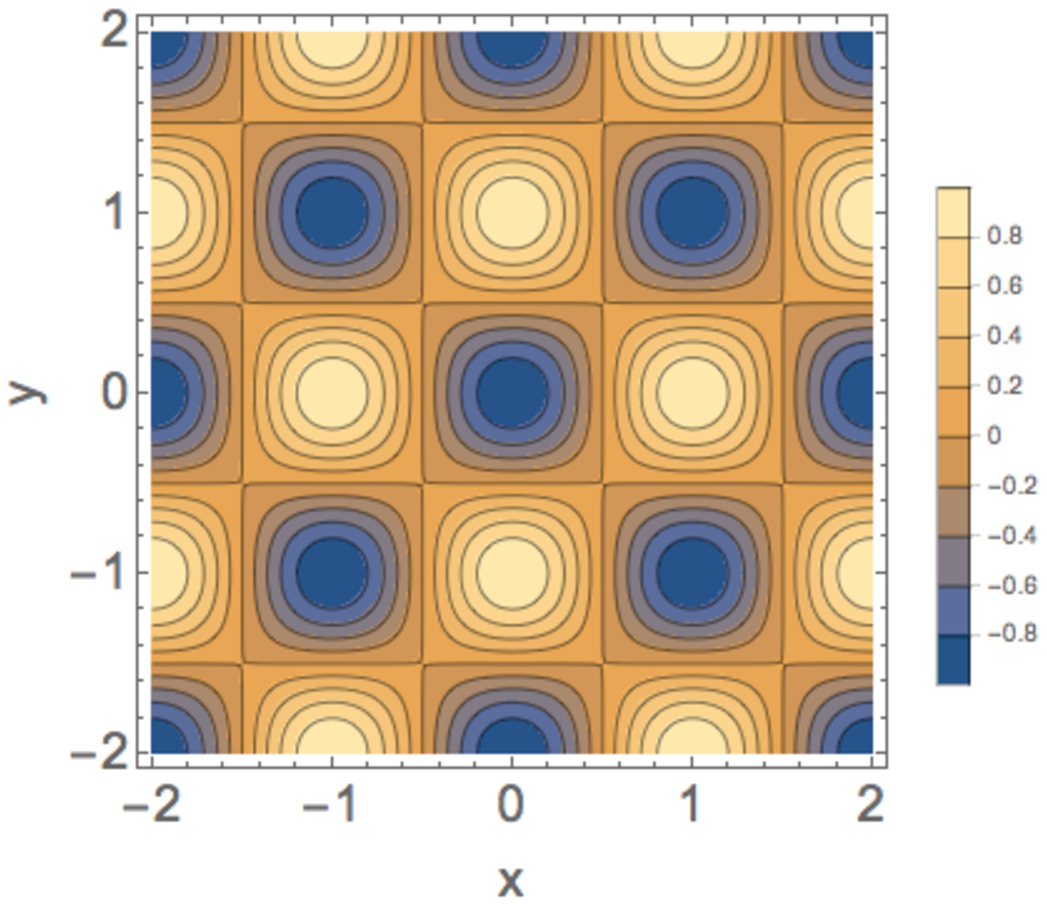}
\end{minipage}
}
{
\begin{minipage}[b]{0.3\textwidth}
\includegraphics[width=1\textwidth]{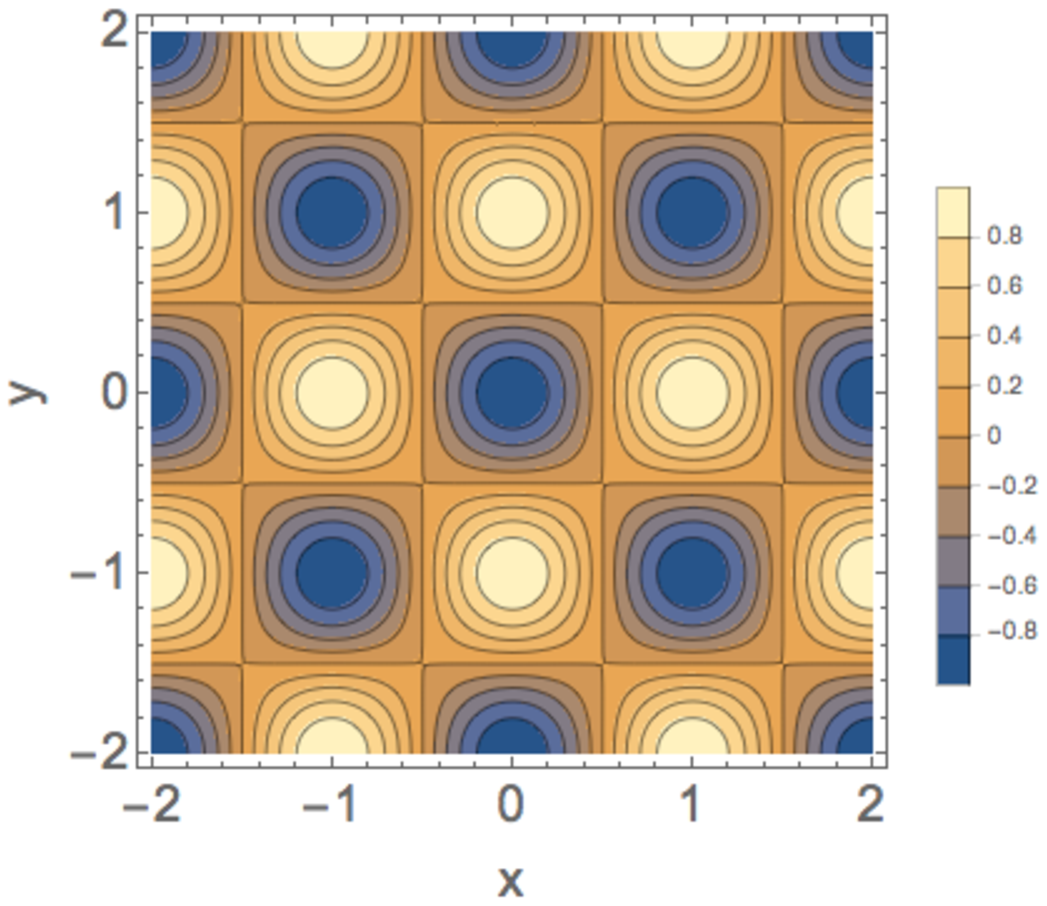}
\end{minipage}
}
{
\begin{minipage}[b]{0.3\textwidth}
\includegraphics[width=1\textwidth]{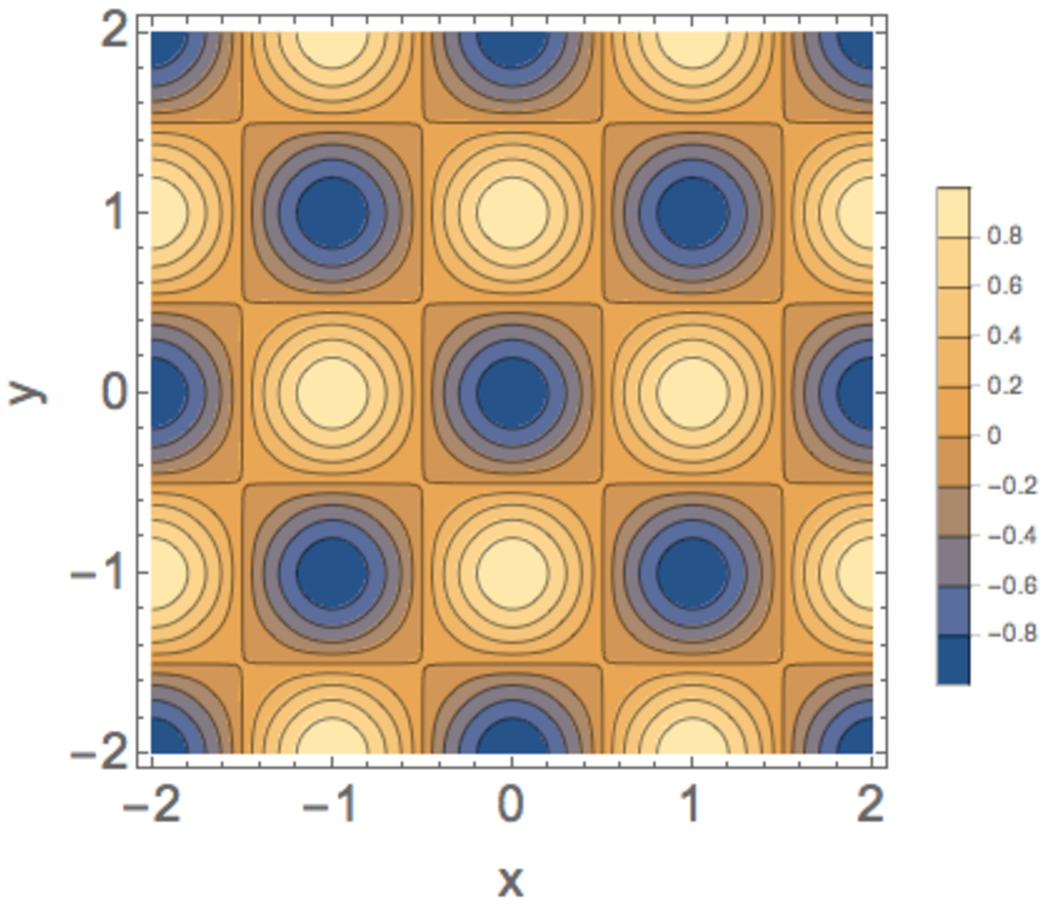}
\end{minipage}
}
\caption[]{The distributions of $\phi_0(x,y,0)$  at $L=1.0~{\rm fm}$ (in half-skyrmion phase) for $\sqrt{eB}=0$ (left),
  $\sqrt{eB}=400~{\rm MeV}$(middle) and $\sqrt{eB}=800~{\rm MeV}$(right).
 }
  \label{dphi0h}
\end{figure*}

\begin{figure*}[!htpb]
\centering
{
\begin{minipage}[b]{0.3\textwidth}
\includegraphics[width=1\textwidth]{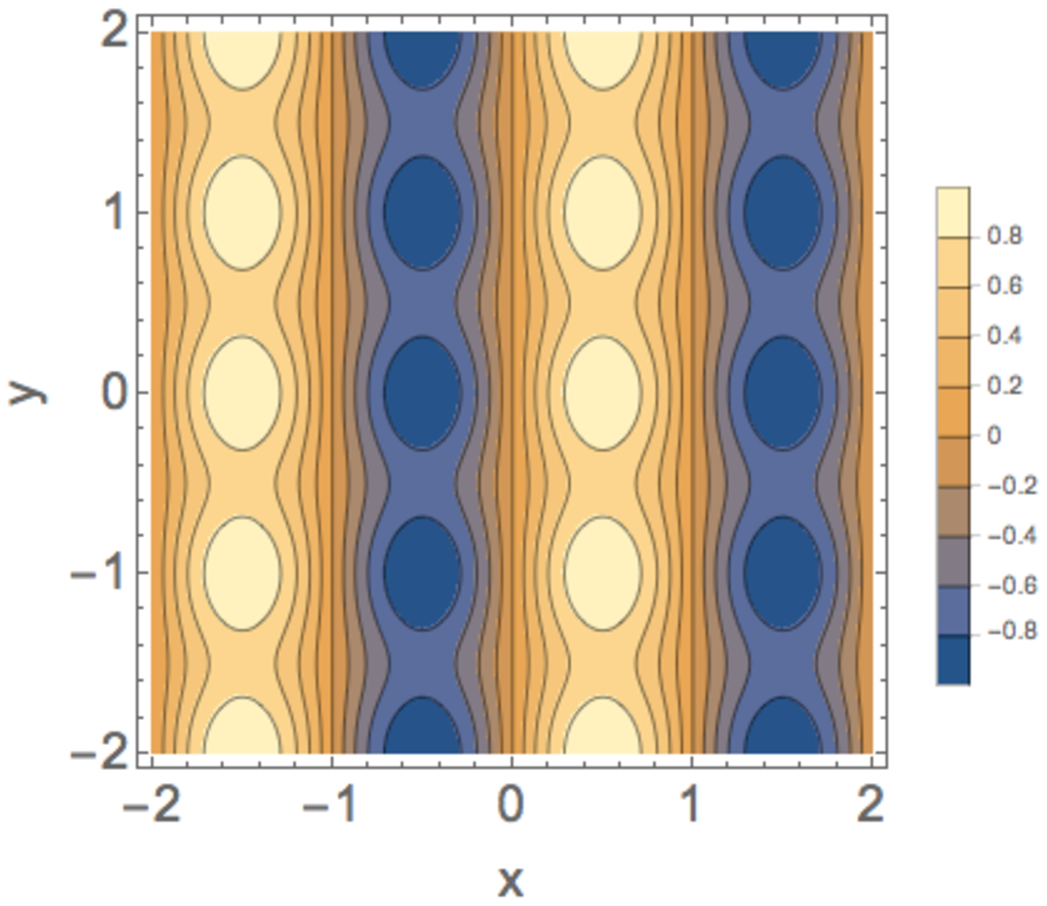}
\end{minipage}
}
{
\begin{minipage}[b]{0.3\textwidth}
\includegraphics[width=1\textwidth]{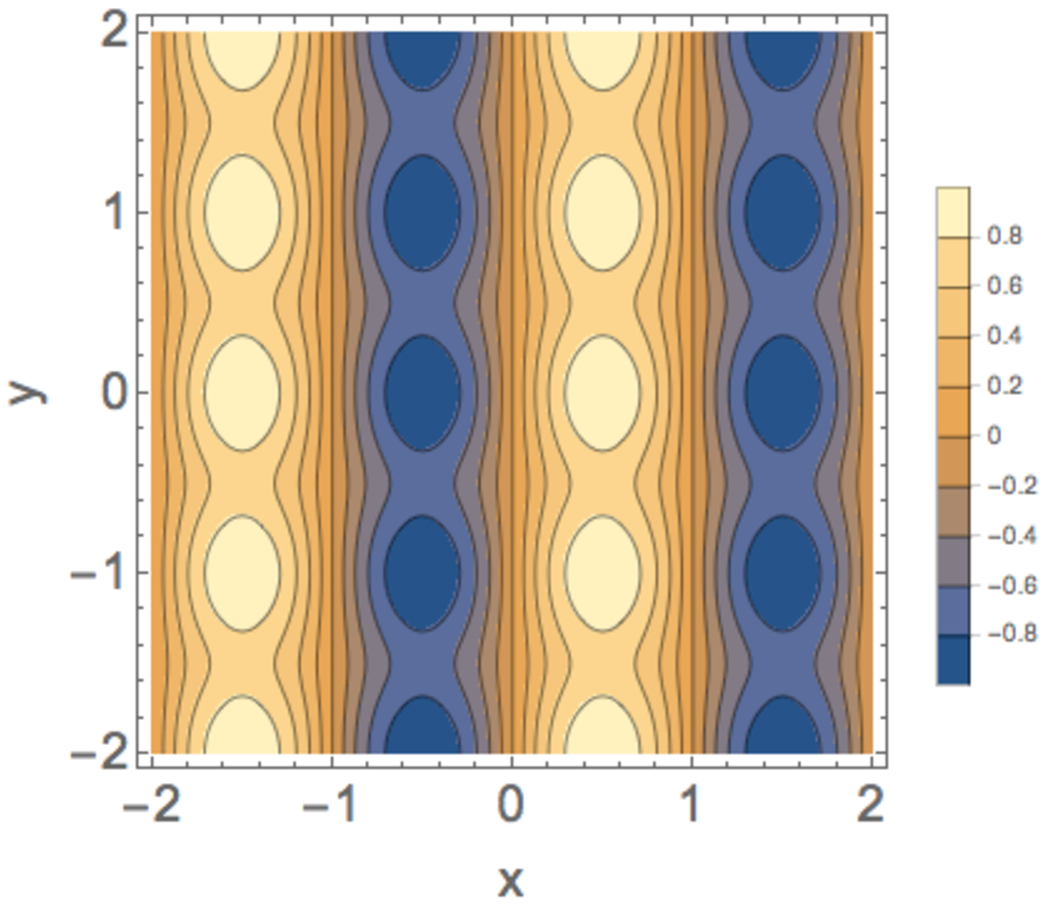}
\end{minipage}
}
{
\begin{minipage}[b]{0.3\textwidth}
\includegraphics[width=1\textwidth]{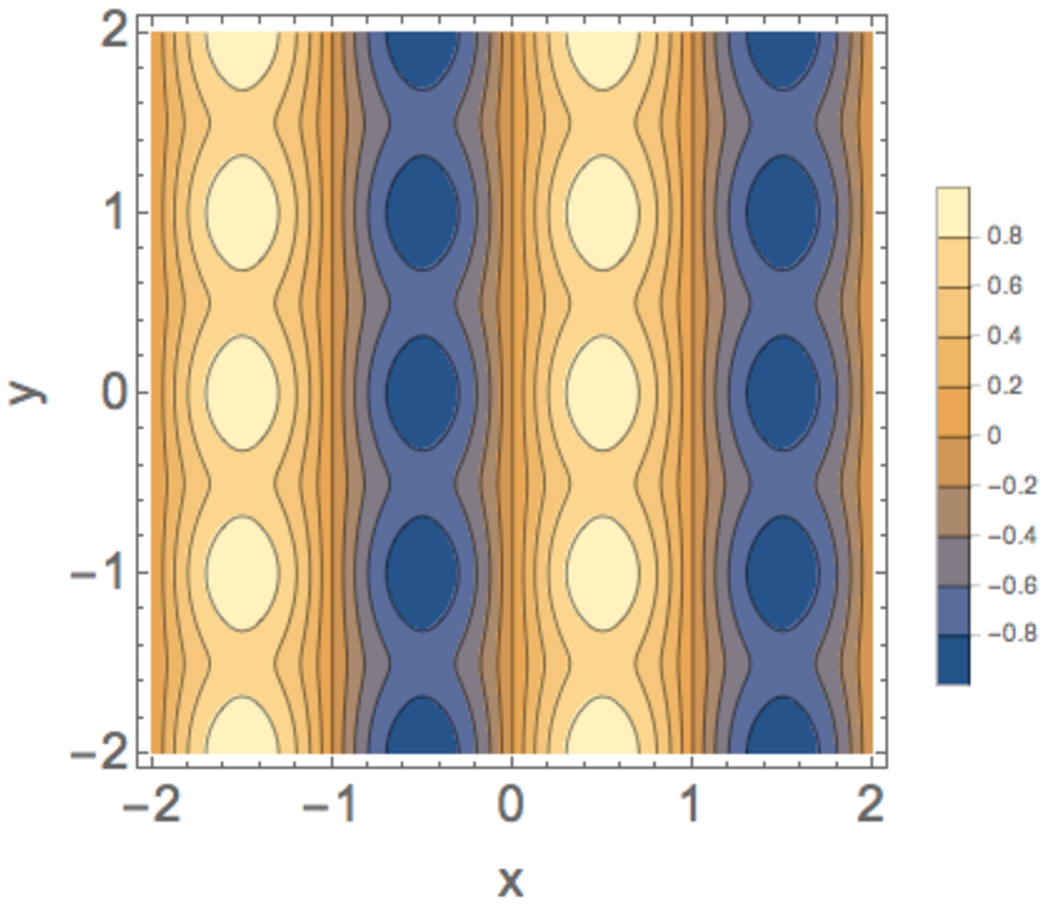}
\end{minipage}
}
\caption[]{The distributions of $\phi_1(x,y,0)$ at $L=1.0~{\rm fm}$ (in half-skyrmion phase) for
 $\sqrt{eB}=0$ (left), $\sqrt{eB}=400~{\rm MeV}$(middle) and $\sqrt{eB}=800~{\rm MeV}$(right).
 }
  \label{dphi1h}
\end{figure*}

\begin{figure*}[!htpb]
\centering
{
\begin{minipage}[b]{0.4\textwidth}
\includegraphics[width=1\textwidth]{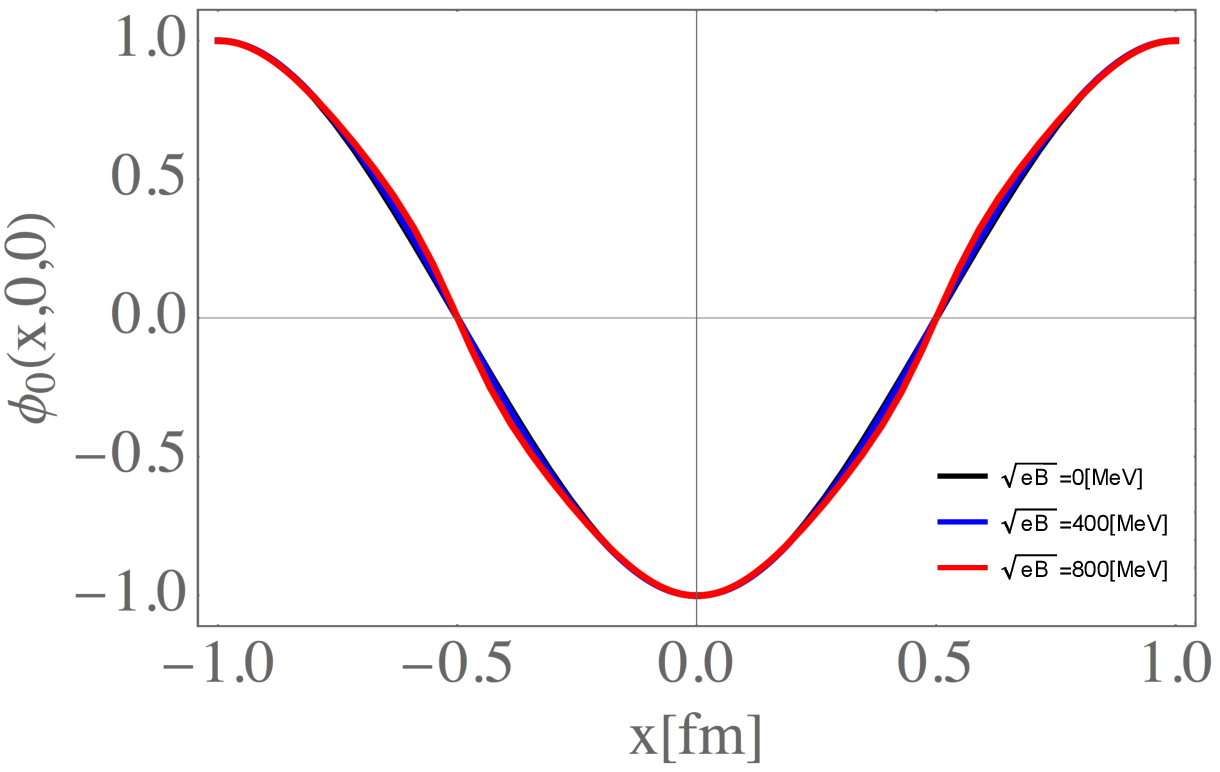}
\end{minipage}
}
{
\begin{minipage}[b]{0.4\textwidth}
\includegraphics[width=1\textwidth]{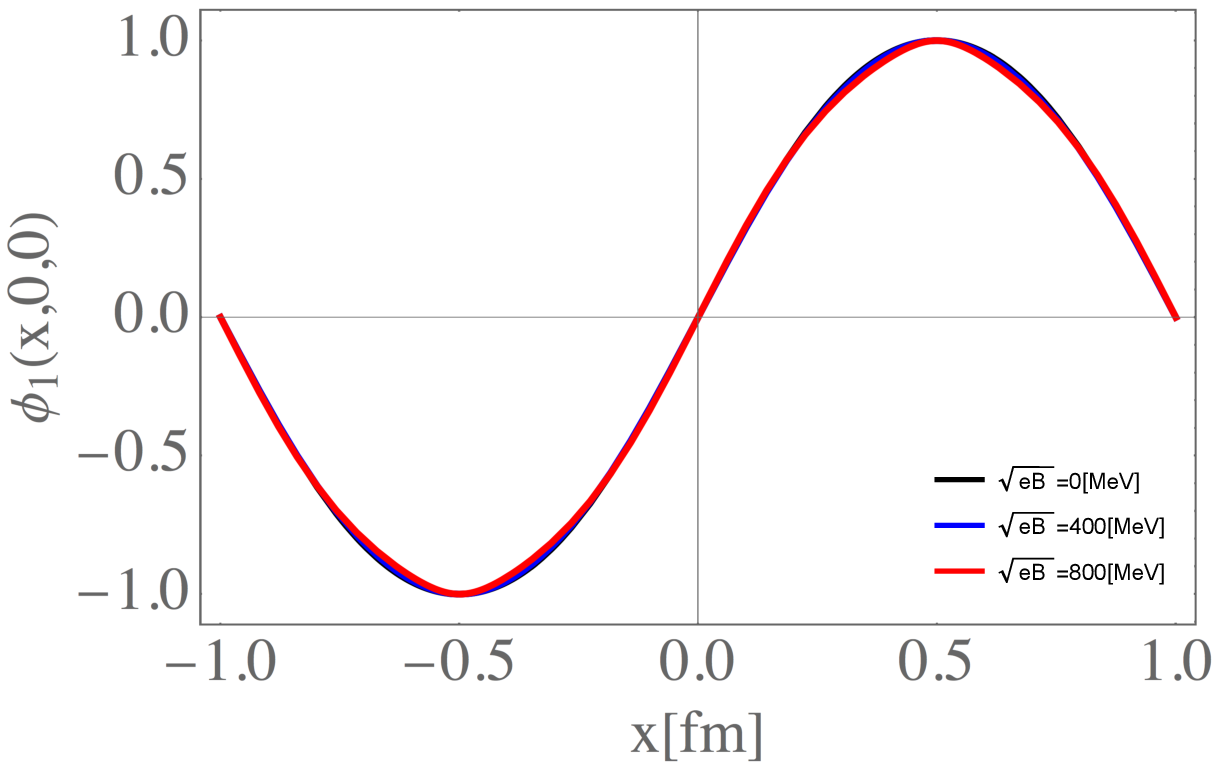}
\end{minipage}
}
\caption[]{The distributions of $\phi_0(x,0,0)$ (left panel) and $\phi_1(x,0,0)$ (right panel)
 at $L=1.0~{\rm fm}~$ (in half-skyrmion phase) with $\sqrt{eB}$ varied.
 }
  \label{dphi-x-HS}
\end{figure*}

We next show in Figs.~\ref{dphi0h} and~\ref{dphi1h} the magnetic dependence on the distributions of $\phi_0(x,y,z)$ and $\phi_1(x,y,z)$ in the half-skyrmion phase at $L=1.0~{\rm fm}$.
In these figures the left panels show the inhomogeneities of $\phi_0$ and $\phi_1$ without the magnetic field.
The result seen from those left panels are, again,
consistent with the analysis in~\cite{Harada:2015lma}.
When the magnetic field is turned on (middle and right panels), one can see that the $\phi_0$ and $\phi_1$ inhomogeneities are hardly affected by the strength of $eB$. This is in contrast to the situation in the skyrmion phase. This can also be illustrated by zooming in the $y=z=0$ plane in Fig.~\ref{dphi-x-HS}.

\subsection{Pion decay constant}
\label{decay_cons}

We next turn to the order parameter of the chiral symmetry breaking, $f_\pi$, in the skyrmion crystal with a magnetic effect. We introduce the fluctuating pion field through
\begin{eqnarray}
U & = & \breve{u} \bar{U} \breve{u}, \nonumber\\
\breve u & = & \exp\left( i\pi^a\tau^a/(2f_\pi)
\right),
\end{eqnarray}
where $\bar{U}$ is the background skyrmion field and $\pi^a$ describes the fluctuating pion field.
Thus, the medium-modified pion decay constant, $f_\pi^*$,  is obtained as~\cite{Lee:2003aq}
\begin{eqnarray}
\frac{f_\pi^*}{f_\pi} & = &
\sqrt{1-\frac{2}{3}(1-\langle\phi_0^2\rangle)}.
\end{eqnarray}

\begin{figure*}[!htpb]
 \begin{center}
   \includegraphics[width=8cm]{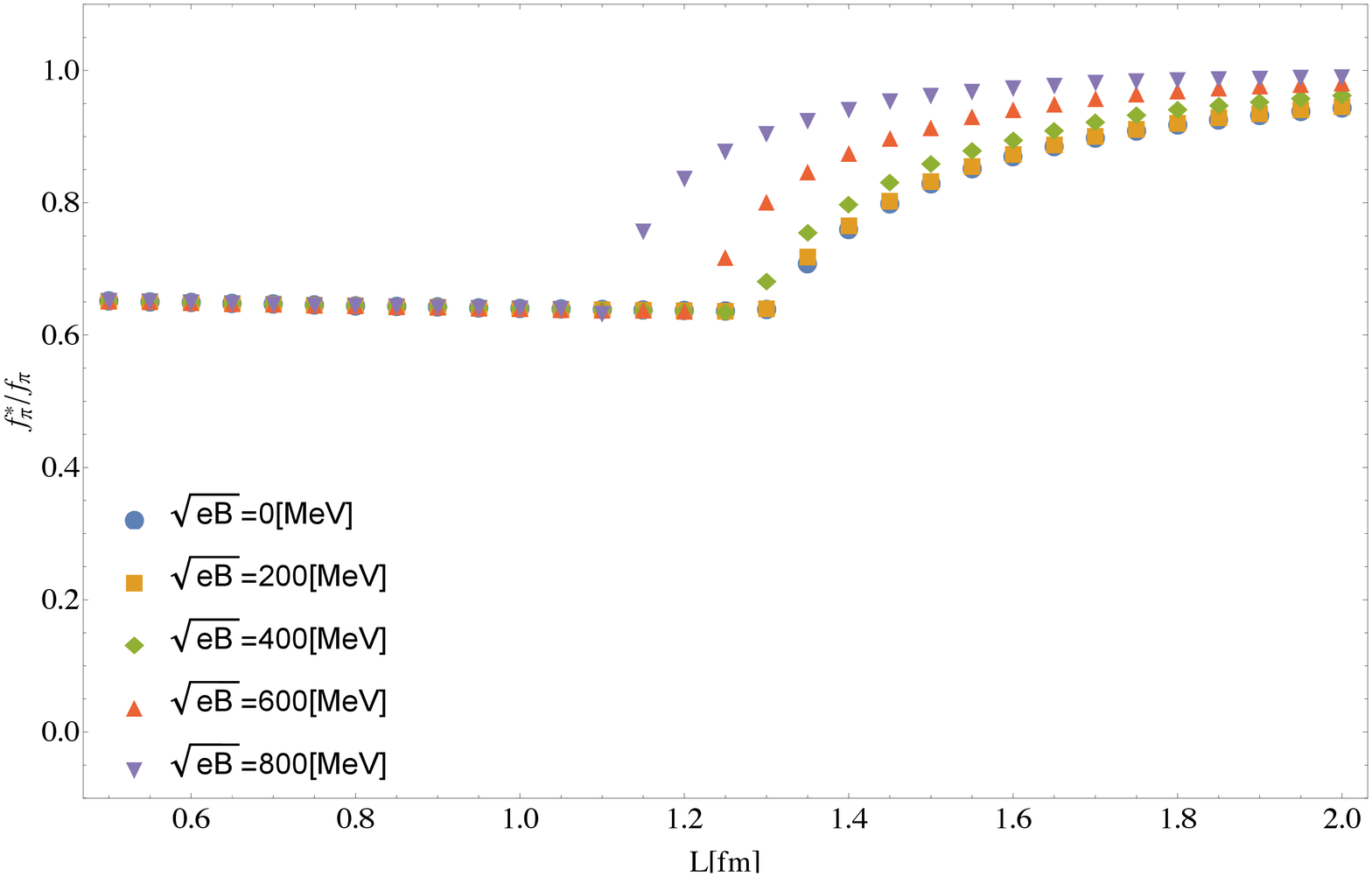}
  \end{center}
 \caption{The pion decay constant normalized to the matter-free value, $f_\pi^*/f_\pi$, as a function of the crystal size $L$ with different choice of the magnetic field. }
\label{fpi_syme}
\end{figure*}

In Fig.~\ref{fpi_syme} we plot $f_\pi^*/f_\pi$ as a function of
the crystal size of $L$ with the magnetic field varied.
The density dependence at ${\sqrt eB}=0$
agrees with the result of~\cite{Lee:2003aq}~\footnote{
$f_\pi^*/f_\pi$ can be vanishing at the chiral phase transition point if
one takes into account a chiral-singlet("dilaton") effect as discussed in
\cite{Park:2003sd}}.
Furthermore, in the presence of magnetic field,
the magnitude of the chiral symmetry breaking gets larger
when the strength is increased.

\subsection{Deformation of the skyrmion configuration}
\label{baryon_density}

We finally in this section discuss the deformation of the skyrmion configuration and the baryon shape in the presence of a magnetic field.

In the skyrmion crystal approach the skyrmion configuration can be extracted by plotting the position dependence of the the baryon-number density distribution functions in (\ref{bar_dens_eq}). From (\ref{bar_dens_eq}) one can check that the winding number density  $\rho_W(x,y,z)$ keeps the crystal symmetries for the FCC and CC structures in the presence of a  magnetic field. However, the induced-baryon number density $\tilde \rho_{eB}$ does not have this feature. This implies that the skyrmion configurations in both phases would be significantly deformed by the presence of the magnetic field.

\begin{figure*}[!htpb]
\centering
{
\begin{minipage}[b]{0.3\textwidth}
\includegraphics[width=1\textwidth]{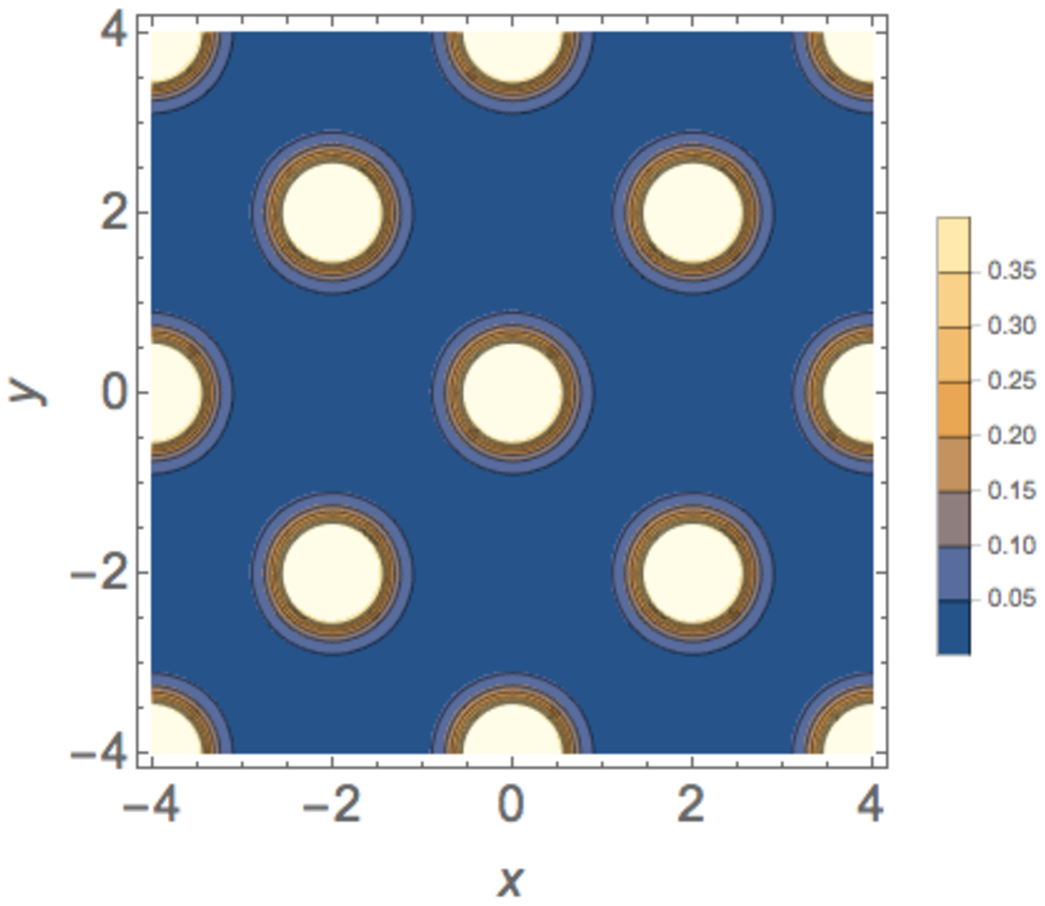}
\end{minipage}
}
{
\begin{minipage}[b]{0.3\textwidth}
\includegraphics[width=1\textwidth]{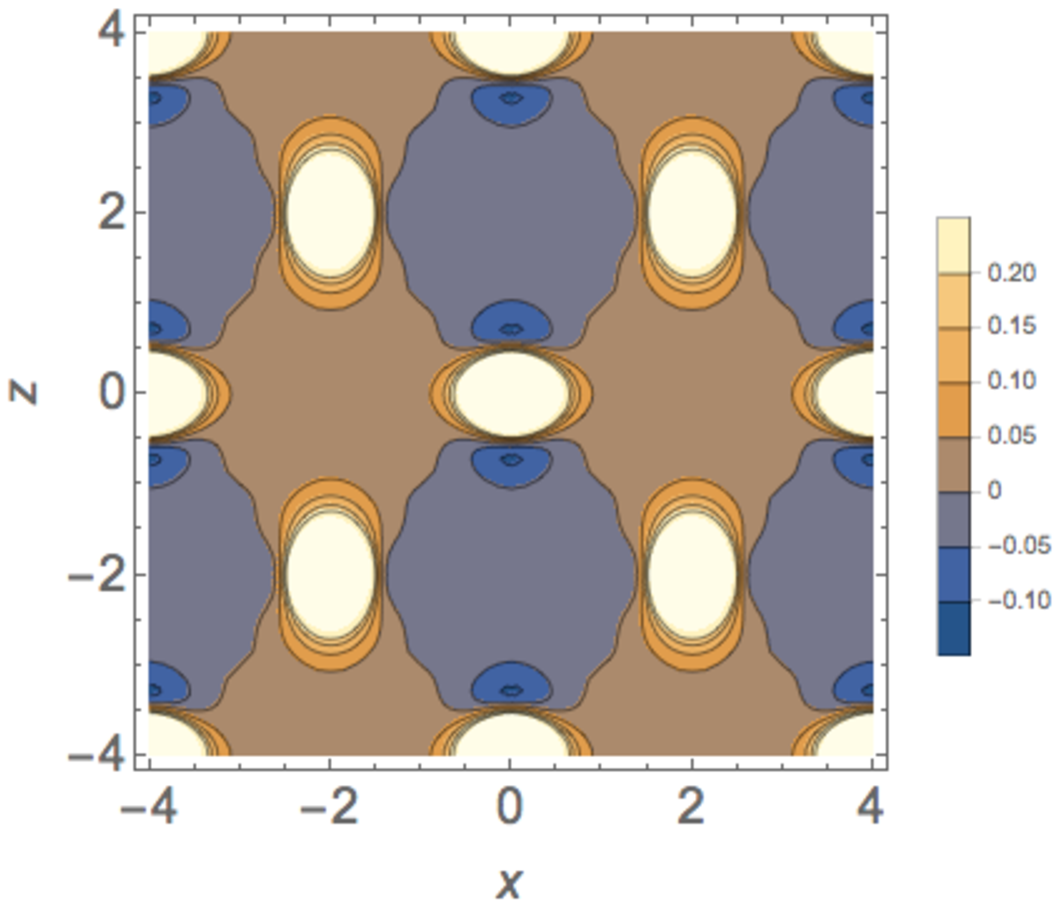}
\end{minipage}
}
{
\begin{minipage}[b]{0.3\textwidth}
\includegraphics[width=1\textwidth]{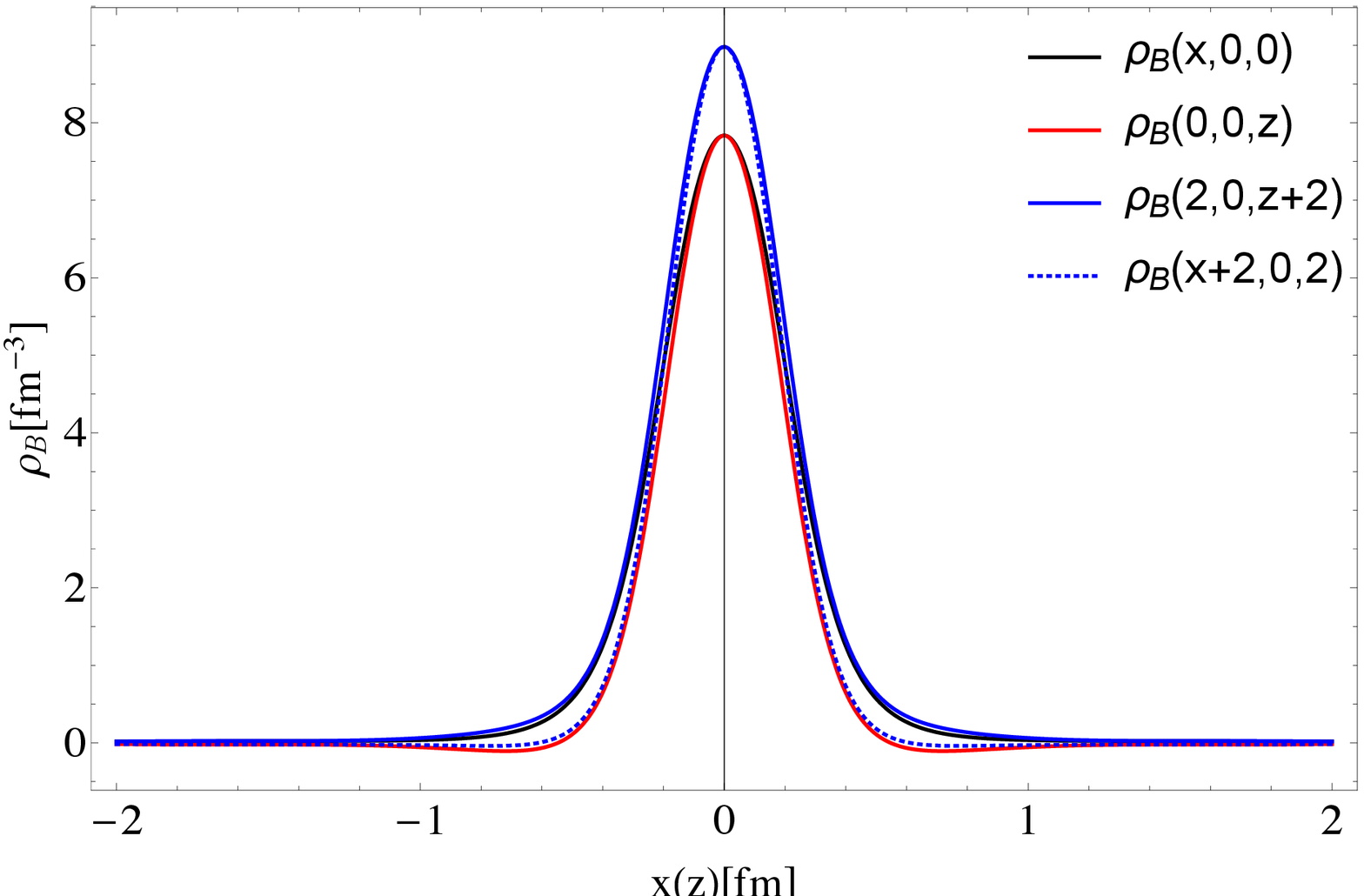}
\end{minipage}
}
\caption[]{The skyrmion configurations at $\sqrt{ eB}=400$~{\rm MeV} and $L=2.0~{\rm fm}$ (in skyrmion phase).
  The left(middle) panel displays the density contour plot on x-y plane(x-z plane), and the right panel corresponds to the distribution along the x-axis  or z-axis including the size rescaled by $L=2.0~{\rm fm}$.
 }
  \label{bd_400_2}
\end{figure*}
\begin{figure*}[!htpb]
\centering
{
\begin{minipage}[b]{0.3\textwidth}
\includegraphics[width=1\textwidth]{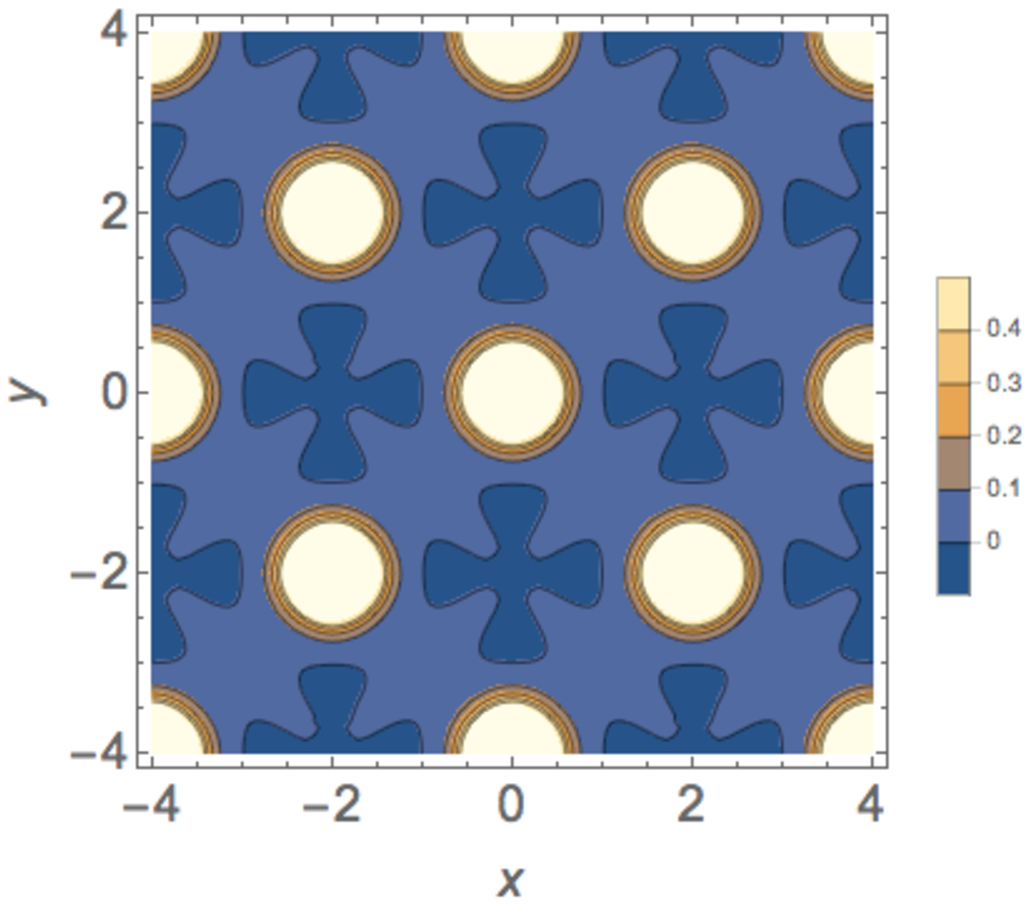}
\end{minipage}
}
{
\begin{minipage}[b]{0.3\textwidth}
\includegraphics[width=1\textwidth]{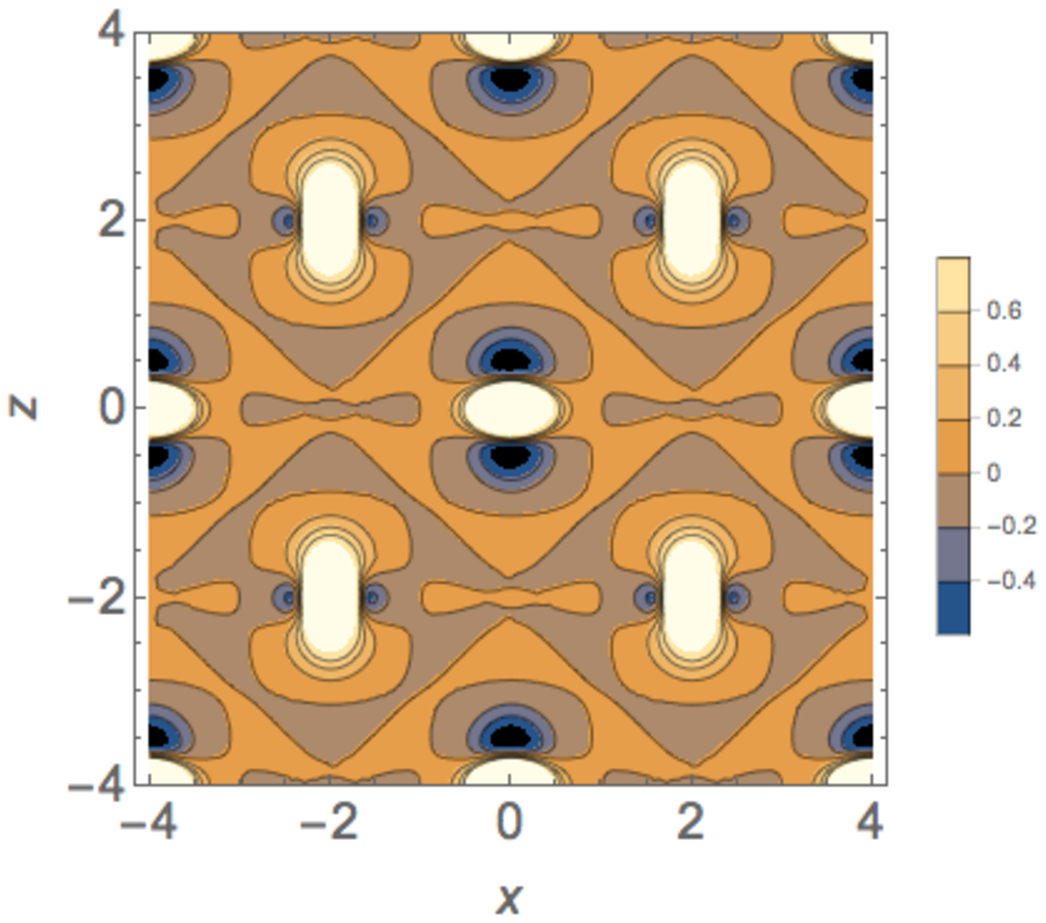}
\end{minipage}
}
{
\begin{minipage}[b]{0.3\textwidth}
\includegraphics[width=1\textwidth]{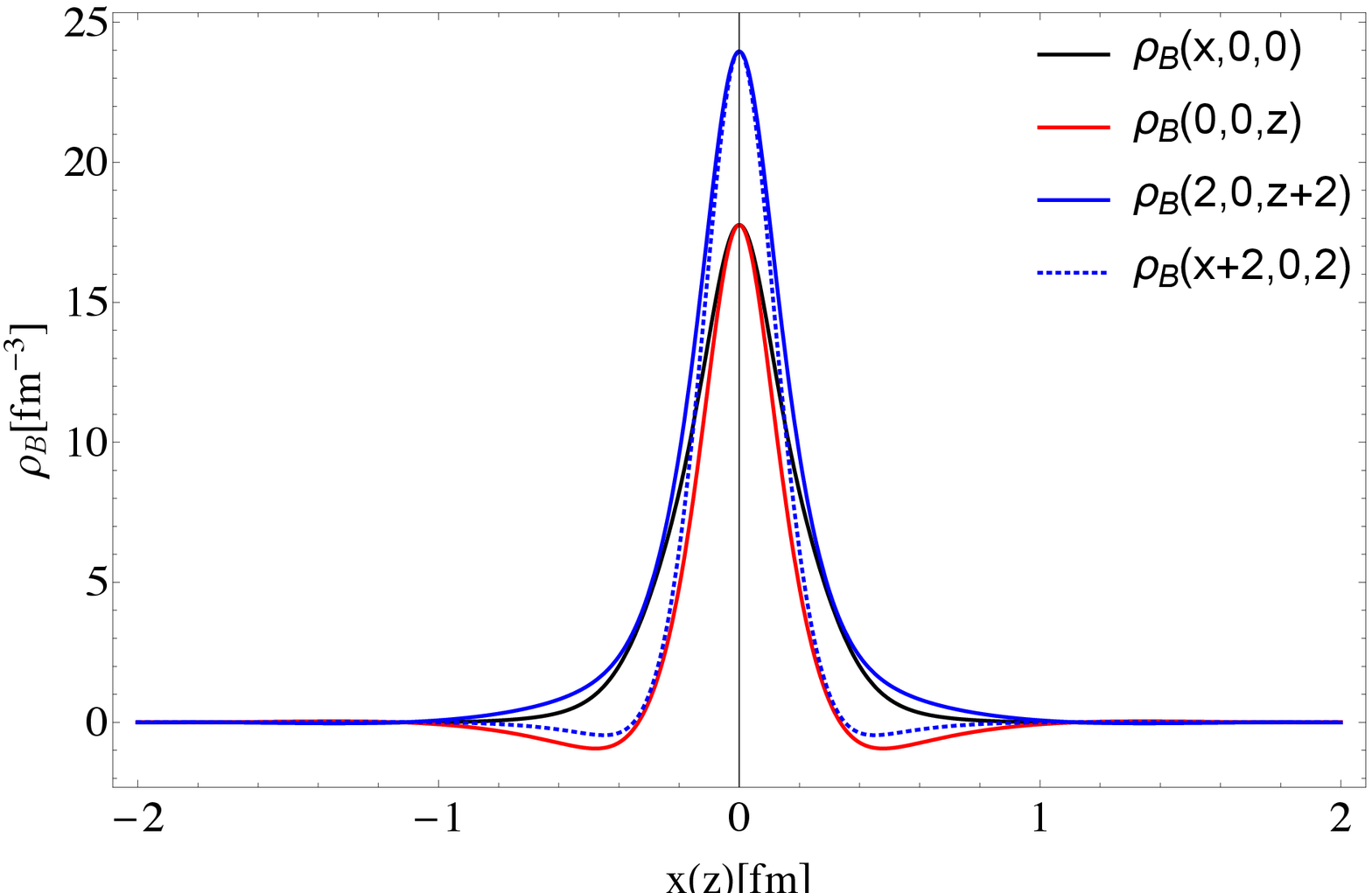}
\end{minipage}
}
\caption[]{ The same as Fig.~\ref{bd_400_2} but with $\sqrt{eB}=800~{\rm MeV}$}.

  \label{bd_800_2}
\end{figure*}

In Figs.~\ref{bd_400_2} and~\ref{bd_800_2}, we plot the skyrmion configurations in the skyrmion phase.
First, it is interesting to note that even for a large magnetic field $eB$, the FCC structure essentially holds
(see, in particular, Fig. \ref{bd_800_2} for $\sqrt{eB}=800~{\rm MeV}$).
For the single baryon shape (corresponding to higher-intense objects in Figs.~\ref{bd_400_2} and~\ref{bd_800_2}),
one also finds that it is deformed to be an elliptic form
by the magnetic field.
The deformation of this kind has also been found in the isolated skyrmion analysis in matter-free space \cite{He:2015zca}.

\begin{figure*}[!htpb]
\centering
{
\begin{minipage}[b]{0.3\textwidth}
\includegraphics[width=1\textwidth]{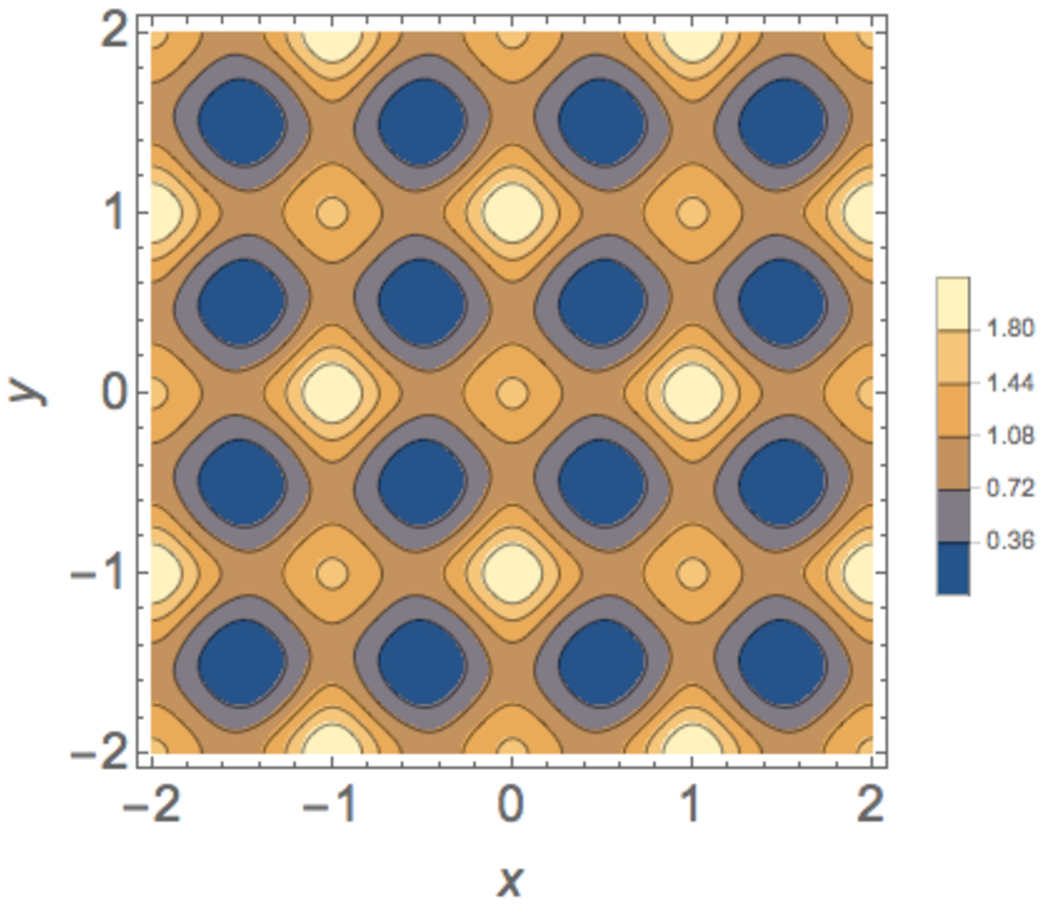}
\end{minipage}
}
{
\begin{minipage}[b]{0.3\textwidth}
\includegraphics[width=1\textwidth]{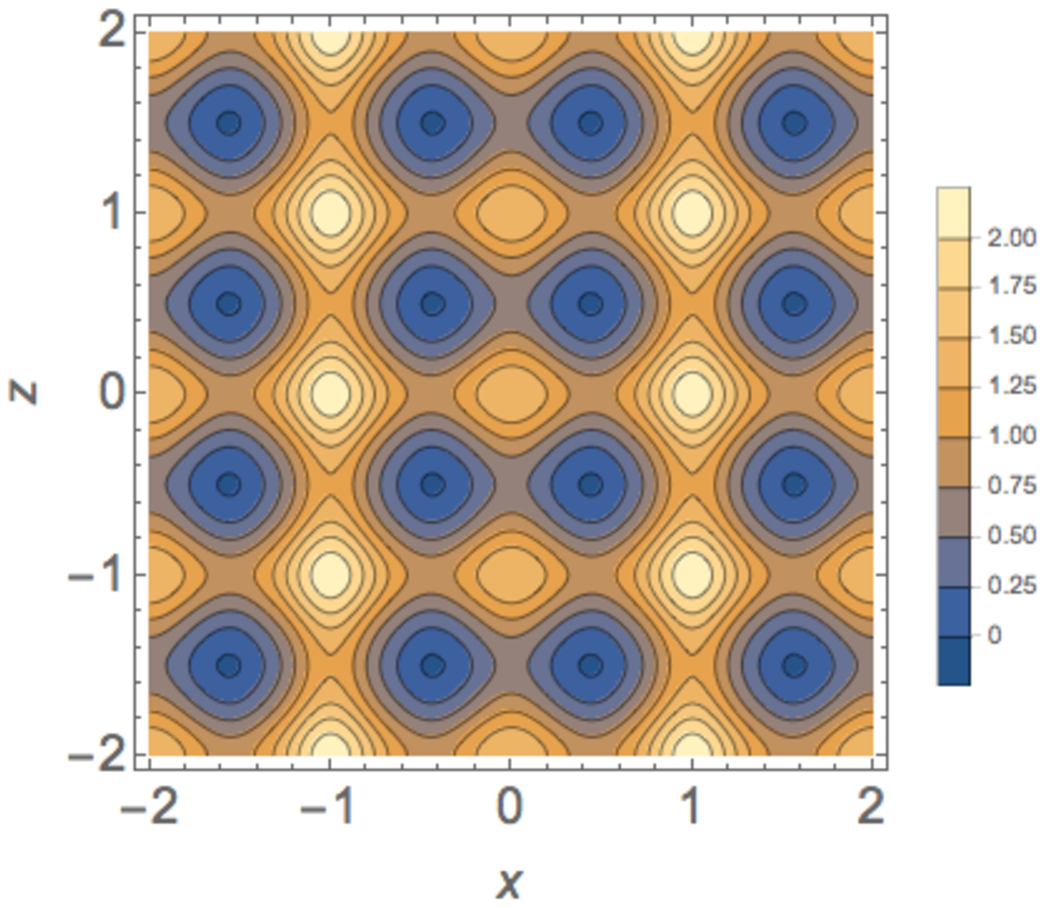}
\end{minipage}
}
{
\begin{minipage}[b]{0.3\textwidth}
\includegraphics[width=1\textwidth]{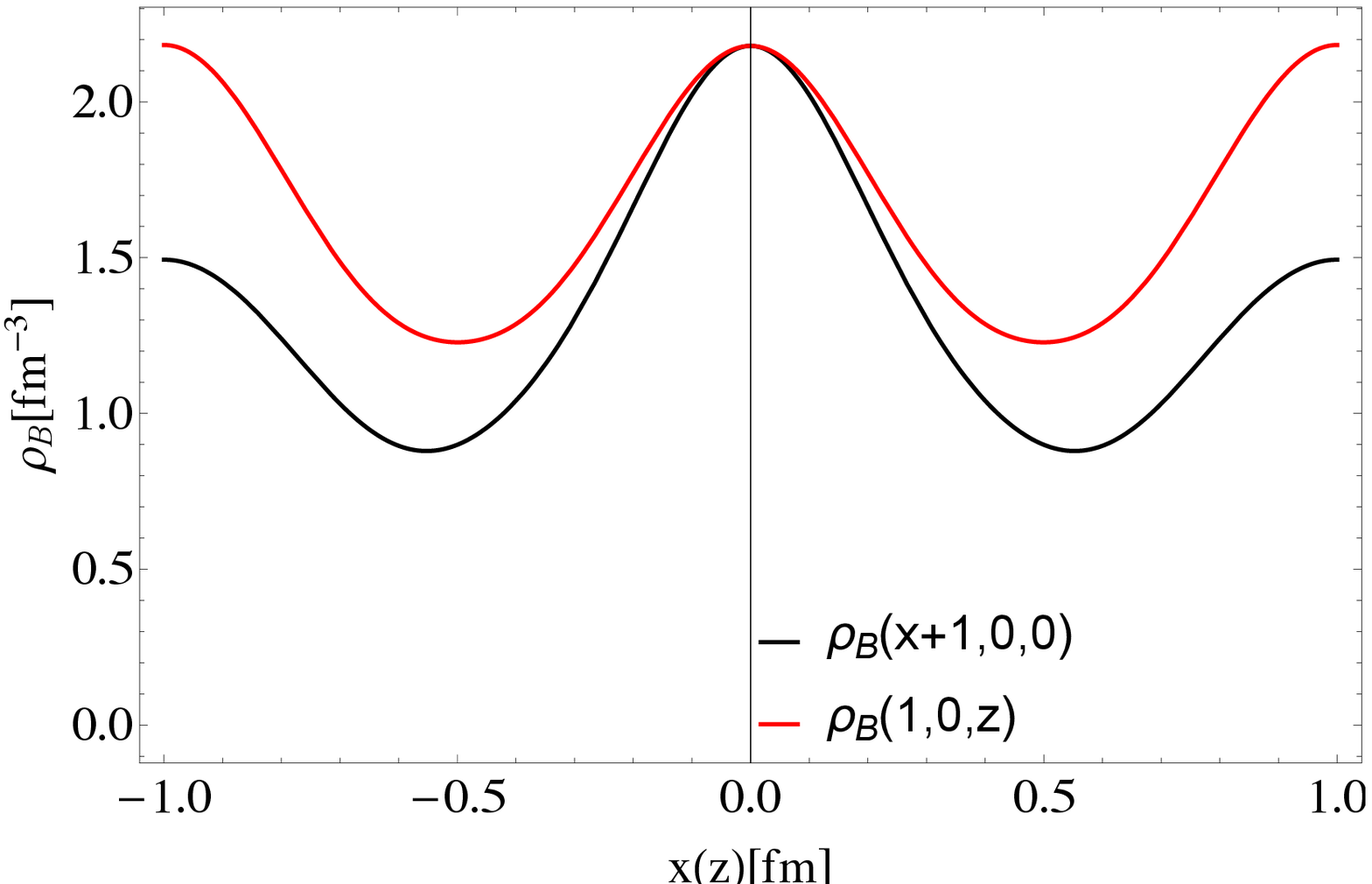}
\end{minipage}
}
\caption[]{ The skyrmion configurations at $\sqrt{ eB}=400$~{\rm MeV} and $L=1.0~{\rm fm}$ (in half-skyrmion phase).
  The left(middle) panel displays the density contour plot on x-y plane(x-z plane), and the right panel corresponds to the distribution along the x-axis or z-axis including the size of shift by $L=1.0~{\rm fm}~$.
 }
  \label{bd_400_1}
\end{figure*}

\begin{figure*}[!htpb]
\centering
{
\begin{minipage}[b]{0.3\textwidth}
\includegraphics[width=1\textwidth]{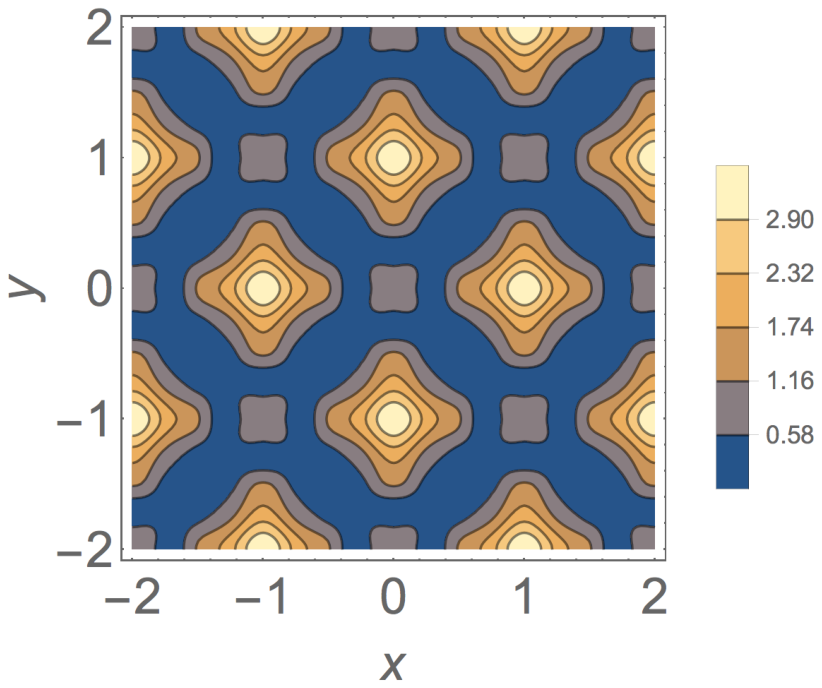}
\end{minipage}
}
{
\begin{minipage}[b]{0.3\textwidth}
\includegraphics[width=1\textwidth]{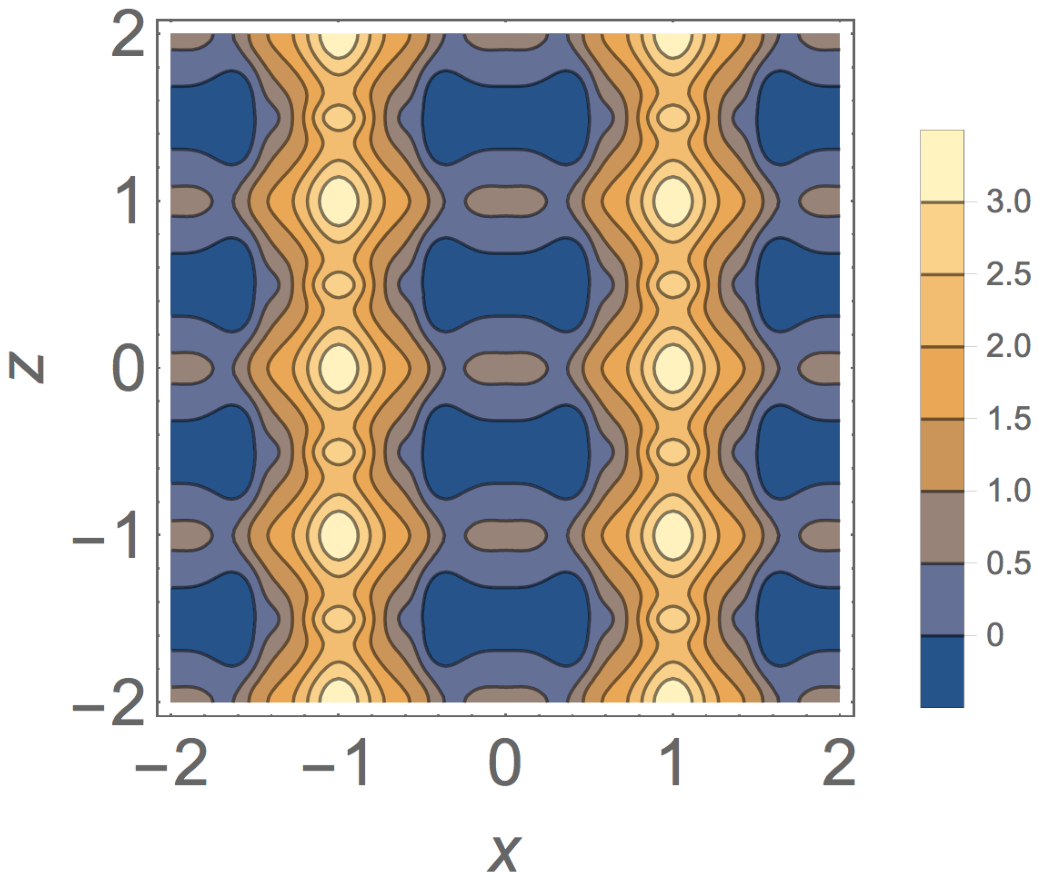}
\end{minipage}
}
{
\begin{minipage}[b]{0.3\textwidth}
\includegraphics[width=1\textwidth]{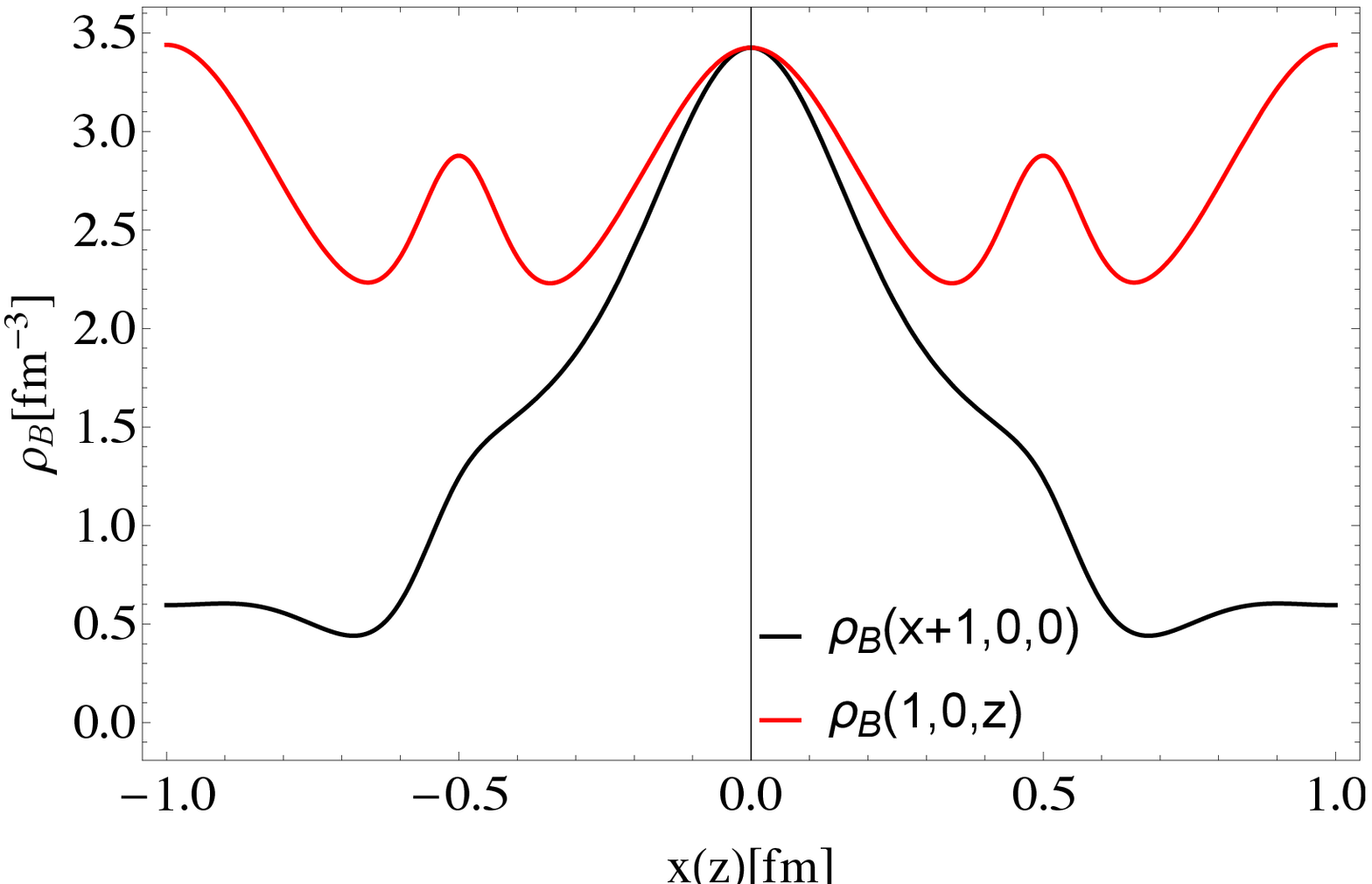}
\end{minipage}
}
\caption[]{ The same as Fig.~\ref{bd_400_1} but with $\sqrt{ eB}=800$~{\rm MeV}.
 }
  \label{bd_800_1}
\end{figure*}

We now move on to the half-skyrmion phase.
We make plots of the half-skyrmion configurations
in Figs.~\ref{bd_400_1} and~\ref{bd_800_1}.
One can immediately see that the half-skyrmion configuration is dramatically deformed by the existence of the magnetic field and the magnetic effect not only breaks the CC form, but also makes a multiple-peak structure (Fig.\ref{bd_800_1}). Those nontrivial deformations in the half-skyrmion phase
would be indirect probes for the presence of inhomogeneous chiral condensate shown in Fig.~\ref{dphi1h}.

\clearpage
\section{Summary}

\label{sec:sum}
In this paper, we analyzed magnetic field effects on the nuclear matter based on
the skrymion crystal approach for the first time. Several interesting phenomena have been found:
\begin{itemize}

\item
The magnetic effect plays the role of a catalyzer for
the topological phase transition (topological deformation for the skyrmion crystal
configuration from the skrymion phase to half-skyrmion phase), and
the per-baryon energy  (soliton mass) is enhanced by
the presence of the magnetic field.

\item
Even in the presence of the magnetic field,
the inhomogeneous chiral condensate created by the crystal structure
persists both in the skyrmion and half-skyrmion phases.
Remarkably, as the strength of magnetic field gets larger,
the inhomogeneous chiral condensate in the skyrmion phase tends to be drastically localized,
while in the half-skyrmion phase the inhomogeneity configuration is hardly affected.

\item
A large magnetic effect in a low density region (in the skyrmion phase) makes the baryon shape to be an elliptic form,
while the crystal configuration is intact.
In contrast, in a high density region (in the half-skyrmion phase),
the crystal structure is dramatically changed due to the existence of strong magnetic field.

\item
A nontrivial deformation of the skrymion configuration due to a large magnetic field
would be a novel indirect probe for the presence of the inhomogeneity
of the chiral condensate in the half-skyrmion phase.

\end{itemize}

In closing, we shall make a few comments on what we have found and 
the related prospect on its phenomenological implications. 
As is well known, 
the skyrmion crystal model is not appropriate 
to simulate the nuclear matter at a low density region, 
such as a region below the normal nuclear density, 
which is almost around the critical point for 
the topological phase transition in the model.  
Though might quantitatively be somewhat away from the realistic situation for 
nuclear matters,  
the magnetic effect on the phase transition such as the magnetic catalysis    
would qualitatively involve curious enough aspects for the high dense matter physics 
in a strong magnetic field. 
 A strong magnetic field might be generated in the core of 
 neutron stars or magnetars in correlation with a chiral dynamics, 
 as was discussed in~\cite{Son:2007ny,Eto:2012qd}.  
In that case, 
the remnant of
the topological phase transition including the characteristic magnetic effect 
may be 
incorporated into the equation of state for the neutron stars, 
as in~\cite{Paeng:2017qvp,Li:2018ykx,Ma:2018jze}, 
to simulate the compact star properties, 
which might be observed in the near future experiments. \\

\acknowledgments

M.~K. would like to thank the support from Jilin University where the early stage of this work was done.
The work of M.~K. is supported in part by JSPS Grant- in-Aid for JSPS Research Fellow No. 18J15329.
 Y.~L. M. was supported in part by National Science Foundation of China (NSFC) under Grant No. 11475071, 11747308 and the Seeds Funding of Jilin University. The work of S.~M. was supported in part by
the JSPS Grant-in-Aid for Young Scientists (B) No. 15K17645.

\appendix

\begin{widetext}

\section{Discretization of $x\partial_i\phi_a$}

In this Appendix, we present the method for discretizing the quantity including a derivative. For example, we consider the quantity $x\partial_x\phi_3$,
\begin{eqnarray}
[x\partial_x\phi_3]_{\rm disc}=
\frac{[x\partial_x \bar\phi_3]_{\rm disc}}{\sqrt{\bar \phi_a \bar\phi_a}}
-\frac{(\bar\phi_b[x\partial_x\bar\phi_b]_{\rm disc})\bar\phi_3}{(\bar\phi_a\bar\phi_a)^{3/2}}.
\label{A1}
\end{eqnarray}
We make discretizations for the square bracket parts denoted as
$[\,\,\,\,]_{\rm disc}$.
The $[x \partial_x \bar{\phi}_0]_{\rm disc}$ part is computed as
\begin{eqnarray}
x\partial_x\bar\phi_0(x,y,z)
&=&x\partial_x
\int_{0}^{\infty} \frac{dp_x}{(2\pi)}\int_{0}^{\infty} \frac{dp_y}{(2\pi)}\int_{0}^{\infty}  \frac{dp_z}{(2\pi)}
\bar\phi_0({\bm p})8
\cos(p_x x)\cos(p_yy)\cos(p_zz)\nonumber\\
&=&
\int_{0}^{\infty} \frac{dp_x}{(2\pi)}\int_{0}^{\infty} \frac{dp_y}{(2\pi)}\int_{0}^{\infty}  \frac{dp_z}{(2\pi)}
\bar\phi_0({\bm p})8
\left[
p_x\partial_{p_x}\cos(p_x x)\right]
\cos(p_yy)
\cos(p_zz)\nonumber\\
&\xrightarrow{{\rm discretization}}&
\sum_{a,b,c}\bar \beta_{abc}
\frac{a\pi}{L}
\frac{\cos\{(a+2)\pi x/L\}-\cos(a\pi x/L)}{2\pi/L}\cos(b\pi y/L)
\cos(c\pi z/L)\nonumber\\
&\equiv&
[x\partial_x\bar\phi_0]_{\rm disc}]_{\rm disc}(x,y,z)
\,.
\end{eqnarray}
Similarly, for other terms, we have
\begin{eqnarray}
{[}x\partial_x\bar \phi_1]_{\rm disc}&=&\sum_{h,k,l}\bar \alpha_{hkl}^{(1)}\frac{h\pi}{L}
\frac{\sin\bigl\{(h+2)\pi x/L\bigl\}-\sin(h\pi x/L)}{2\pi/L}
\cos(k\pi y/L)\cos(l\pi z/L) \,, \nonumber\\
{[}x\partial_x\bar \phi_2]_{\rm disc}&=&\sum_{h,k,l}\bar \alpha_{hkl}^{(2)}
\frac{l\pi}{L}
\frac{\cos\bigl\{(l+2)\pi x/L\bigl\}-\cos(l\pi x/L)}{2\pi/L}
\sin(h\pi y/L)\cos(k\pi z/L) \,, \nonumber\\
{[}x\partial_x\bar \phi_3]_{\rm disc}&=&\sum_{h,k,l}\bar \alpha_{hkl}^{(3)}
\frac{k\pi}{L}
\frac{\cos\bigl\{(k+2)\pi x/L\bigl\}-\cos(k\pi x/L)}{2\pi/L}
\cos(l\pi y/L)\sin(h\pi z/L)
\,.
\end{eqnarray}
Putting those terms into the right-hand side of Eq.(\ref{A1}), we thus obtain the discretized form
of $x \partial_x \phi_3$.
\end{widetext}


\end{document}